\documentclass[10.75pt, a4paper]{article}
\usepackage{titlesec}
\titleformat{\section}
  {\bf\sffamily}
  {\thesection. }
  {5pt}
  {\MakeUppercase}
\renewcommand{\thesection}{\Roman{section}} 

\titleformat{\subsection}
  {\bf\sffamily}
  {\thesubsection. }
  {5pt}{}
  
\renewcommand{\thesubsection}{\Alph{subsection}}

\usepackage[affil-it]{authblk} 
\usepackage{etoolbox}
\usepackage{lmodern}

\usepackage{eurosym}

\usepackage[utf8]{inputenc}
\usepackage{geometry}
\usepackage[hidelinks]{hyperref}
\usepackage[citestyle=numeric,style=phys,biblabel=brackets,backend=bibtex,sorting=none,url=false,doi=false, natbib]{biblatex}
\bibliography{bibliography}
\DefineBibliographyStrings{english}{andothers={\itshape et\addabbrvspace al\adddot}}

\usepackage{csquotes}
\usepackage[]{authblk} 
\usepackage{etoolbox}
\usepackage{lmodern}

\usepackage{amsmath}
\usepackage{enumerate}
\usepackage{enumitem}
\usepackage{graphicx}
\usepackage{siunitx}
\usepackage{float}
\usepackage[]{parskip} %skip=0.7em, indent=1.5em

\makeatletter
\patchcmd{\@maketitle}{\LARGE \@title}{\fontsize{16}{19.2}\selectfont\@title}{}{}
\makeatother

\usepackage[normalem]{ulem}

\usepackage{graphicx}% Include figure files
\usepackage{dcolumn}% Align table columns on decimal point
\usepackage{bm}% bold math
\usepackage{tikz}
\usetikzlibrary{math}
\usetikzlibrary{shapes.geometric, arrows}
\usetikzlibrary{positioning}
\usetikzlibrary{shapes,arrows}
\usetikzlibrary{intersections}
\usepackage{csvsimple}
\usepackage{changepage}
\usepackage[tbtags]{mathtools}
\usepackage{siunitx}
\usepackage{csquotes}

\usepackage{float}
\usepackage{subfigure}
\usepackage[
	nonumberlist, 				% keine Seitenzahlen anzeigen
	acronym,      				% ein Abk�rzungsverzeichnis erstellen
	nomain,						% no main glossary
	nopostdot,					% Kein punkt nach Beschreibung
]{glossaries}
\usepackage{glossary-superragged}
\usepackage[hidelinks]{hyperref}
\usepackage{color, colortbl}

\usepackage{wrapfig}

\usepackage{pgfplots}
\usepackage{tikz}
\usetikzlibrary{arrows}
\usepackage{amsmath}
\pgfplotsset{compat=newest}
\usepgfplotslibrary{fillbetween}
\usetikzlibrary{calc}
\def\centerarc[#1](#2)(#3:#4:#5)% Syntax: [draw options] (center) (initial angle:final angle:radius)
{ \draw[#1] ($(#2)+({#5*cos(#3)},{#5*sin(#3)})$) arc (#3:#4:#5); }

\usepackage{amssymb}

\usepackage{nicefrac}

\usepackage{abstract}
\usepackage{multirow}
\usepackage{tabularx}
\usepackage{booktabs}
\usepackage{array}
\usepackage{longtable}
\newcolumntype{L}[1]{>{\raggedright\let\newline\\\arraybackslash\hspace{0pt}}m{#1}}
\newcolumntype{C}[1]{>{\centering\let\newline\\\arraybackslash\hspace{0pt}}m{#1}}
\newcolumntype{R}[1]{>{\raggedleft\let\newline\\\arraybackslash\hspace{0pt}}m{#1}}

\newacronym{3d}{3D}{three dimensional}
\newacronym{am}{AM}{additive manufacturing}
\newacronym{fdm}{FDM}{fused deposition modeling}
\newacronym{ism}{ISM}{in-space manufacturing}
\newacronym{iss}{ISS}{International Space Station}
\newacronym{fcb}{FCB}{Functional Cargo Block}
\newacronym{dem}{DEM}{discrete element method}
\newacronym{md}{MD}{molecular dynamics}
\newacronym{dc}{DC}{direct-current}
\newacronym[plural=PFCs,firstplural=parabolic flight campaigns (PFCs)]{pfc}{PFC}{Parabolic Flight Campaign}
\newacronym{fft}{FFT}{Fast Fourrier Transform}
\newacronym{cad}{CAD}{Computer Assisted Design}
\newacronym{ptfe}{PTFE}{polytetrafluoroethylene}
\newacronym{ps}{PS}{polystyrene}
\newacronym{nasa}{NASA}{National Aeronautics and Space Administration}
\newacronym{esamm}{ESAMM}{Extended Structure Additive Manufacturing Machine}
\newacronym{amf}{AMF}{Additive Manufacturing Facility}
\newacronym{us}{US}{United States}
\newacronym{usa}{USA}{United States of America}
\newacronym{bmgs}{BMGs}{Bulk Metallic Glasses}
\newacronym{esa}{ESA}{European Space Agency}
\newacronym{si}{SI}{International System of Units, abbreviated from French \textit{Syst\`{e}me International (d'unit\'{e}s)}}
\newacronym{dlr}{DLR}{German Aerospace Center}
\newacronym{liggghts}{LIGGGHTS}{\acrshort{lammps} Improved for General Granular and Granular Heat Transfer Simulations}
\newacronym{lammps}{LAMMPS}{Large-scale Atomic/Molecular Massively Parallel Simulator}
\newacronym{sjkr}{SJKR}{Simplified Johnson-Kendall-Roberts}
\newacronym{ded}{DED}{Directed Energy Deposition}
\newacronym{slm}{SLM}{Selective Laser Melting}
\newacronym{sls}{SLS}{Selective Laser Sintering}
\newacronym{eva}{EVA}{Extra-Vehicular Activity}
\newacronym{sem}{SEM}{Scanning Electron Microscopy}
\newacronym{RPM}{RPM}{Ramdom Positioning Machine}
\newacronym{rpm}{rpm}{revolutions per minute}
\newacronym{rise}{RISE}{Research Internships in Science and Engineering}
\newacronym{daad}{DAAD}{German Academic Exchange Service, abbreviated from German \textit{Deutscher Akademischer Austauschdienst}}
\newacronym{fsm}{FSM}{finite-state machine}
\newacronym{ir}{IR}{infrared}
\newacronym{pcbs}{PCBs}{Printed Circuit Boards}
\newacronym{pcb}{PCB}{Printed Circuit Board}
\newacronym{mcr}{MCR}{Modular Compact Rheometer}
\newacronym{sff}{SFF}{Solid Freeform Fabrication}
\newacronym{uv}{UV}{ultraviolet}
\newacronym{abs}{ABS}{acrylonitrile butadiene styrene}
\newacronym{hpde}{HPDE}{high density polyethylene}
\newacronym{pei}{PEI}{polyetherimide}
\newacronym{bff}{BFF}{BioFabrication Facility}
\newacronym{lens}{LENS}{Laser Engineered Net Shaping}
\newacronym{cnc}{CNC}{Computer Numerical Control}
\newacronym{ebf3}{EBF$^3$}{Electron Beam Free-Form Fabrication}
\newacronym{leo}{LEO}{Low Earth Orbit}
\newacronym{pc}{PC}{polycarbonate}
\newacronym{crissp}{CRISSP}{Customisable Recyclable International Space Station Packaging}
\newacronym{Athena}{Athena}{Advanced Telescope for High-ENergy Astrophysics}
\newacronym{lbm}{LBM}{Laser Beam Melting}
\newacronym{bam}{BAM}{Federal Institute for Materials Research and Testing, abbreviated from German \textit{Bundesanstalt f\"{u}r Materialforschung und-pr\"{u}fung}}
\newacronym{pbf}{PBF}{powder bed fusion}
\newacronym{eb}{EB}{Electron Beam}
\newacronym{2d}{2D}{two dimensional}
\newacronym{4d}{4D}{four dimensional}
\newacronym{ft4}{FT4}{Freeman Technology 4 Powder Rheometer}
\newacronym{dsc}{DSC}{Differential Scanning Calorimetry}
\newacronym{pmma}{PMMA}{polymethylmethacrylate}
\newacronym{1g}{$1g$}{gravity on-ground}
\newacronym{mug}{$\mu g$}{microgravity}
\newacronym{bcm}{BCM}{Box Counting Method}
\newacronym{mct}{MCT}{Mode Coupling Theory}
\newacronym{gmct}{gMCT}{granular Mode Coupling Theory}
\newacronym{itt}{ITT}{Integration Through Transients}
\newacronym{mfc}{MFC}{Mass Flow Controller}
\newacronym{ct}{CT}{computed tomography}
\newacronym{xct}{XCT}{X-ray computed tomography}
\newacronym{cv}{CV}{curriculum vitae}
\newacronym{pi}{PI}{principal investigator}
\newacronym{osp}{OSP}{orthogonal superimposed perturbation}
\newacronym{npi}{NPI}{Network Partnering Initiative}
\newacronym{ecsat}{ECSAT}{European Centre for Space Applications and Telecommunications}
\newacronym{eac}{EAC}{European Astronaut Centre}
\newacronym{estec}{ESTEC}{European Space Research and Technology Centre}
\newacronym{fps}{fps}{frames per second}
\newacronym{pdf}{pdf}{probability density function}
\newacronym{al}{Al}{aluminium}
\newacronym{ss}{\textit{SS}}{\textit{Smooth Surface}}
\newacronym{rs}{\textit{RS}}{\textit{Rough Surface}}
\newacronym{rcp}{rcp}{random close packing}
\newacronym{iop}{IoP UvA}{Institute of Physics of the University of Amsterdam}
\newacronym{mp}{MP}{Institute of Material Physics for Space}
\newacronym{elgra}{ELGRA}{European Low Gravity Research Association}
\newacronym{zarm}{ZARM}{Center of Applied Space Technology and Microgravity}
\newacronym{piv}{PIV}{particle image velocimetry}
\usepackage[framemethod=tikz]{mdframed}
\usepackage{lipsum}
\usepackage{dirtytalk}
\usepackage{physics}
\usepackage{tcolorbox}
\usepackage[font=footnotesize]{caption}
\newtcolorbox{mybox}[1]{colback=green!6!white,colframe=black!75!black,fonttitle=\bfseries,title=#1}
\newtcolorbox{mybox2}{colback=red!5!white,colframe=red!75!black}

\usepackage{pifont}

\usepackage{soul,xcolor}
\setstcolor{red}

\usepackage{xcolor,hyperref}
\hypersetup{
   colorlinks,
   linkcolor={blue!50!black},%{red!80!black},
   citecolor={blue!50!black},
   urlcolor={blue!80!black}
} 
\newcommand{\J}{\mathcal{J}} 
\definecolor{mycolor}{rgb}{0.122, 0.435, 0.698}

\title{Coalescence of Printed Yield Stress Filaments in Direct Ink Writing}

\author[1]{Hugo L. Fran\c ca\footnote{h.l.franca@uva.nl, ORCID: 0000-0002-5361-7704}}
\author[2]{Daniël Tieman}
\author[3]{James D. Shemilt\footnote{ORCID: 0000-0002-9158-0930}}
\author[4]{Cassio M. Oishi\footnote{ORCID: 0000-0002-0904-6561}}
\author[1,5]{Maziyar Jalaal\footnote{m.jalaal@uva.nl, mj547@cam.ac.uk, ORCID: 0000-0002-5654-8505}}

\affil[1]{Van der Waals-Zeeman Institute, Institute of Physics, University of Amsterdam, \protect\\
Science Park 904, Amsterdam, 1098XH, The Netherlands}
\affil[2]{Department of Biomedical Engineering and Physics, Amsterdam UMC Location AMC, University of Amsterdam, Meibergdreef 9, 1105AZ Amsterdam, The Netherlands}
\affil[3]{Department of Mathematics, University of British Columbia, Vancouver, BC, V6T 1Z2, Canada}
\affil[4]{Departamento de Matem\'atica e Computa\c c\~ao, Faculdade de Ci\^encias e Tecnologia, \protect\\
Universidade Estadual Paulista ``J\'ulio de Mesquita Filho'', Presidente Prudente, Brazil}
\affil[5]{Department of Applied Mathematics and Theoretical Physics, \protect\\
University of Cambridge, Wilberforce Road, Cambridge CB3 0WA, UK}

\begin{document}
\definecolor{brickred}{rgb}{0.8, 0.25, 0.33}
\definecolor{darkorange}{rgb}{1.0, 0.55, 0.0}
\definecolor{persiangreen}{rgb}{0.0, 0.65, 0.58}
\definecolor{persianindigo}{rgb}{0.2, 0.07, 0.48}
\definecolor{cadet}{rgb}{0.33, 0.41, 0.47}
\definecolor{turquoisegreen}{rgb}{0.63, 0.84, 0.71}
\definecolor{sandybrown}{rgb}{0.96, 0.64, 0.38}
\definecolor{blueblue}{rgb}{0.0, 0.2, 0.6}
\definecolor{ballblue}{rgb}{0.13, 0.67, 0.8}
\definecolor{greengreen}{rgb}{0.0, 0.5, 0.0}
\begingroup
\sffamily
\date{}
\maketitle
\endgroup

\begin{abstract}
In direct ink writing (DIW), neighbouring filaments of yield-stress inks are deposited side-by-side and are expected to merge into smooth, mechanically robust structures. Unlike Newtonian filaments, coalescence can arrest in finite time, leaving a permanent, non-flat ridge set by the competition between capillarity and rheology. Here we study the coalescence of two printed yield-stress filaments, combining scaling theory for the arrested state, direct numerical simulations, and DIW experiments on Carbopol gels imaged by optical coherence tomography. In the viscoplastic limit, we predict and observe an approximately linear decrease of the final bridge height with plastocapillary number and a critical yield stress above which coalescence does not initiate. Simulations further show that elasticity becomes important at high plastocapillary number, enabling larger final bridge heights via a crossover from a rigid Herschel--Bulkley solid to a deformable Kelvin--Voigt response. Our findings provide a framework for predicting deposition profiles and, ultimately, for mitigating residual topography in DIW.

\textbf{Yield Stress Fluids $|$ Plastocapillarity $|$ Viscoplastic Fluids $|$ Elasto-Viscoplasticity  $|$ Coalescence $|$ 3D Printing $|$ Direct Ink Writing} 

\end{abstract}

% \pagebreak

\section{Introduction}
Droplet coalescence is a fundamental process in fluid mechanics and underpins many natural phenomena and technologies, from the coalescence of raindrops in clouds to inkjet printing. Consequently, the coalescence of Newtonian liquid droplets, both free drops in air and sessile droplets on solid substrates have been studied extensively; see Eggers \emph{et al.}~\cite{Eggers2025} for a recent review. 
Nevertheless, many modern applications involve non-Newtonian fluids. In particular, manufacturing techniques like 3D (bio-)printing rely on complex inks that possess elasto-viscoplastic properties~\cite{saadi2022direct,wei2023go,ho2025direct,sauret2026fluid}. Hence, the effects of both viscoelasticity and yield stress on coalescence have received attention in recent years. Viscoelasticity in polymeric liquids was shown to potentially alter the dynamics of droplet coalescence, particularly in early times~\cite{Varma2022,Dekker2022,Kaneelil2026,Rostami2025}. In viscoelastic solids, elasticity has also been used as a mechanism to arrest coalescence of droplets, with the final shape being modeled through a balance between capillary and elastic forces which, respectively, drive and resist the process \cite{Pawar2012, Dahiya2016}.  
%Yield stress, on the other hand, makes coalescence fundamentally different by potentially arresting the merging process.
Another mechanism of arresting coalescence of droplets is through yield-stress. 
Yield-stress fluids behave as viscoelastic solids when the stress is below a threshold (yield stress), and thus two merging drops of such materials may stop coalescing once the capillary-driven stresses fall below this threshold value. Kern et al.~\cite{Kern2022} observed that two sessile yield-stress droplets initially form a liquid bridge that grows in time much like a viscous Newtonian bridge, but then the growth halts at a finite height and the droplets reach an equilibrium fused shape connected by a \say{frozen} bridge. 
This arrested coalescence of yield stress fluids at a finite time with a non-trivial permanent shape is determined by the balance of surface tension and yield stress (and potentially gravity). Similar plastocapillary phenomena – where capillarity is opposed by a yield stress – have been documented in the spreading of individual viscoplastic drops on surfaces \cite{Jalaal2021,Jalaal2015,Jalaal2018,martouzet2021dynamic,Franca2024,heitmeier2025spreading}. In those cases, the droplet only spreads to a finite radius and final height once stresses everywhere fall below the yield limit. 

Such incomplete relaxation is often undesirable in additive manufacturing applications. A prominent example is Direct Ink Writing (DIW), a widely used extrusion-based 3D printing technique in which viscoplastic filaments are deposited in a layer-by-layer fashion to construct multilayered structures~\cite{saadi2022direct,colanges20232,jang2021effect,yoon2025minimizing,sauret2026fluid}. In this context, the yield stress of the ink plays a dual role: it is essential for shape retention, allowing freshly printed filaments to maintain their form without spreading or collapsing; yet it also poses challenges when adjacent filaments are expected to coalesce. Inadequate merging between neighboring lines can leave behind residual ridges, seams, or voids, which in turn degrade surface smoothness, dimensional accuracy, and interlayer adhesion. These issues complicate the precise translation of digital designs into physical objects, often requiring tedious parameter tuning and process optimization for each ink-printer combination to achieve acceptable fidelity and mechanical integrity.

Much related to DIW, and the present work, van der Kolk \emph{et al.}~ \cite{vanderKolk2023} studied the deposition and spreading of a single yield-stress filament on a substrate, providing insight into how a viscoplastic filament expands to a final width under surface tension. In this study, we go beyond a single filament, and investigate the coalescence of two filaments of a yield-stress fluid from theoretical, computational, and experimental perspectives. To this end, we consider a simplified 2D droplet coalescence geometry (as a model for two neighboring printed lines) and analyze how the interplay of surface tension and rheology dictates the merging dynamics and the final arrested shape. By developing scaling arguments and simulations, we predict the temporal evolution of the liquid bridge (also called neck in some DIW literature) between the filaments and its final equilibrium profile. We also perform laboratory experiments with a printer and use an optical technique (Optical Coherence Tomography) to observe coalesced filaments. We compare these experiments with the theoretical and computational results and discuss the differences.

\section{Problem Statement: Experiments \& Theoretical Setup}
\begin{figure}[h!]
\centering
\includegraphics[width=1\textwidth]{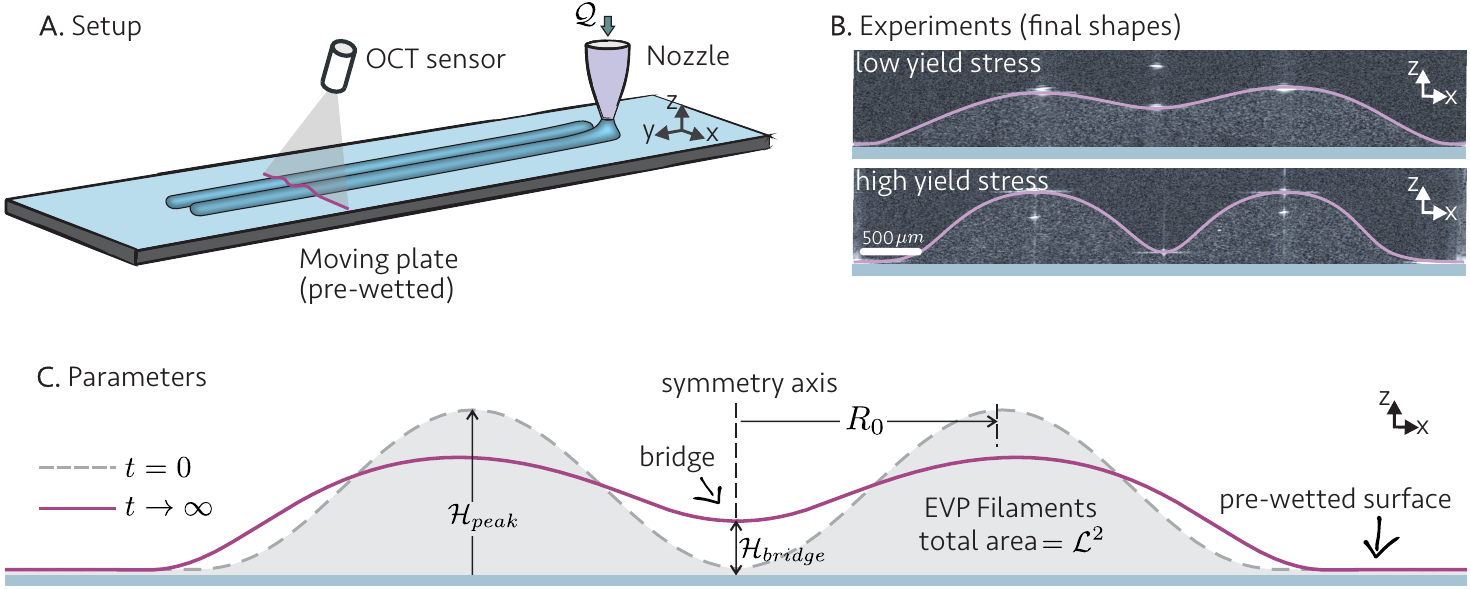}
\caption{Coalescence of two printed yield stress filaments:
\textbf{A}. Experimental setup: two lines of yield–stress fluid are deposited onto a pre-wetted substrate using a custom direct-ink-writing system. A syringe pump extrudes material through a nozzle at a constant flow rate $Q$, while the substrate translates beneath it. The resulting filaments are imaged using Optical Coherence Tomography (OCT). 
\textbf{B}. OCT cross-sections of the coalesced filaments for materials with low and high yield stress, showing the final arrested shapes. The overlaid lines denote the tracked surface profiles. 
\textbf{C}. The coalescence geometry and parameters. Two filaments with combined cross-sectional area $\mathcal{L}^2$, initially separated by a gap, spread and merge to form a central bridge. The bridge height $\mathcal{H}(t)$ evolves until flow ceases due to yielding, producing a well-defined final shape as $t \to \infty$.
}
\label{fig:problem}
\end{figure}

Consider two lines of a yield-stress fluid deposited onto a thin film of the same material on a flat substrate (see Figure~\ref{fig:problem}). Experimentally, we create this configuration using a simple custom-made direct ink writing setup~\cite{vanderKolk2023}. A syringe pump extrudes the material through a nozzle at a constant flow rate $Q$. The nozzle position is fixed in the $x$–$y$ plane but can be manually adjusted in the $z$ direction. The pre-wetted substrate underneath moves in two dimensions via automated linear stages (Thorlabs LTS300/M), which can be programmed.

For each experiment, the surface is first pre-wetted immediately prior to printing to minimize evaporation. The first line is then printed, followed by the second line at a prescribed distance from the first. We wait for 30 seconds to ensure that the filaments have completely spread and reached their final shapes; evaporation in this period is negligible. We use Optical Coherence Tomography (OCT)~\cite{huang1991optical,bouma2022optical} to visualize the interface of the coalesced filaments, with particular emphasis on the final shape (see Figure~\ref{fig:problem}B). Details of the measurement technique can be found in previous work~\cite{vanderKolk2023}, and the material properties, including rheology, are provided in Appendix~\ref{section:app:rheo}.

Our theoretical and computational formulations closely follow the experimental procedure, but with some differences. We consider two filaments, with a combined cross-sectional area of $\mathcal{L}^2$, that begin spreading from a given initial separation at $t = 0$. As they spread, they merge and form a bridge whose height $\mathcal{H}_{\mathrm{bridge}}$ grows in time. Eventually, the flow ceases at a finite time due to the yield stress. Thus, at $t \to \infty$, the system attains a well-defined final shape (Figure~\ref{fig:problem}C). In what follows, we develop theoretical and computational approaches to predict this final shape that is a unique feature of materials with yield stress and cannot be found in Newtonian fluids or viscoelastic polymeric liquids. 

% \begin{figure}[htbp!]
% \centering
% \includegraphics[width=1\textwidth]{Figures/f1.png}
% \caption{\textbf{left:} final bridge height for different $\mathcal{J}$. Solid black circles are computational results. The dashed lines  \textbf{Right:} second invariant of the deformation tensor.}
% \label{fig:f1}
% \end{figure}

\section{Formulations \& Scaling Laws}
\label{sec:formulations}

We numerically approach this problem by solving the full incompressible Navier-Stokes equations. The rheology of the printed filaments is modeled through the elastoviscoplastic constitutive equation proposed by \citet{Saramito2009}, which combines viscoelasticity with a finite yield stress. 

We nondimensionalize all variables using characteristic scales based on the filament geometry and the capillary timescale. The characteristic length is defined as 
\(\mathcal{L}\), where \(\mathcal{L}^2\) is the combined cross-sectional area of the two filaments, while the characteristic velocity and time (for this capillary-driven problem) are respectively given by
\begin{equation}
U_c = \sqrt{\frac{\sigma}{\rho \mathcal{L}}}, \qquad
\tau_c = \sqrt{\frac{\rho \mathcal{L}^3}{\sigma}},
\label{eq:vel_time_scale}
\end{equation}
with \(\rho\) and \(\sigma\) denoting the fluid density and surface tension of the filaments, respectively.  
The nondimensional variables are then introduced as
\begin{equation}
x = \mathcal{L} \, \bar{x}, \quad
\mathbf{u} = U_c \, \bar{\mathbf{u}}, \quad
t = \tau_c \, \bar{t}, \quad
\dot{\boldsymbol{\gamma}} = \frac{U_c}{\mathcal{L}} \, \bar{\dot{\boldsymbol{\gamma}}}, \quad
p = \frac{\sigma}{\mathcal{L}} \, \bar{p}, \quad
\boldsymbol{\tau}^p = \frac{\sigma}{\mathcal{L}} \, \bar{\boldsymbol{\tau}}^p, \quad
\kappa = \frac{1}{\mathcal{L}} \, \bar{\kappa}, \quad
\delta_s = \frac{1}{\mathcal{L}} \, \bar{\delta}_s,
\label{eq:scale_choices}
\end{equation}
where $\mathbf{u}$ is the fluid velocity, $t$ is time, $\dot{\boldsymbol{\gamma}}$ the strain-rate tensor, $p$ the pressure, $\boldsymbol{\tau}^p$ the polymeric stress tensor, $\kappa$ the interfacial curvature and $\delta_s$ a Dirac delta localized on the interface of the filaments that has normal vector $\textbf{n}$. Throughout this paper, all variables shown with a bar on top are considered to be nondimensional according to the rescaling defined above.

With the rescaling choices from equations \eqref{eq:scale_choices}, the full system of equations reads
\begin{align}
&\bar{\rho} \ \frac{D \bar{\mathbf{u}}}{D \bar{t}} 
= - \nabla \bar{p}
+ \nabla \cdot \left( \bar{\eta} \ \mathrm{Oh}_s \, \bar{\dot{\boldsymbol{\gamma}}} \right)
+ \nabla \cdot \bar{\boldsymbol{\tau}}^p
+ \bar{\kappa} \, \bar{\delta}_s \, \mathbf{n}, \label{eq:momentum}\\[3pt]
&\nabla \cdot \bar{\mathbf{u}} = 0, \label{eq:continuity} \\
& \mathrm{De} \, \overset{\nabla}{\bar{\boldsymbol{\tau}}^p}
+ \max\!\left[0, 
\left(\frac{\lVert   \bar{\boldsymbol{\tau}_d}   \rVert - \mathcal{J}}
{\mathrm{Oh}_p^{\,1-n} \lVert   \bar{\boldsymbol{\tau}_d}   \rVert}\right)^{1/n}\right]
\bar{\boldsymbol{\tau}}^p
 = \mathrm{Oh}_p \, \bar{\dot{\boldsymbol{\gamma}}},
\label{eq:constitutive}
\end{align}
where $\mathrm{Oh}_s$ and $\mathrm{Oh}_p$ are respectively the solvent and polymeric Ohnesorge numbers, $\mathrm{De}$ is the Deborah number, and $\mathcal{J}$ is the plastocapillary number, given respectively by
\begin{equation}
\mathrm{Oh}_s = \frac{\eta_s}{\sqrt{\rho \sigma \mathcal{L}}}, \qquad
\mathrm{Oh}_p = \frac{\eta_p}{\sqrt{\rho \sigma \mathcal{L}}}, \qquad
\mathrm{De} = \frac{\eta_p U_c}{G \mathcal{L}}, \qquad
\mathcal{J} = \frac{\tau_y \mathcal{L}}{\sigma},
\end{equation}
where $\eta_s$ and $\eta_p = K\cdot \tau_c^{1 - n}$ are the solvent and polymeric viscosities of the filaments, $\rho$ their density, $\sigma$ surface tension, $G$ the elastic modulus, $\tau_y$ the yield stress, $K$ the Herschel-Bulkley consistency index and $n$ the power index. The viscosity and density functions $\bar{\rho}$ and $\bar{\eta}$ are defined later in this section, in equation~\eqref{eq:rho_mu}. The $\overset{\nabla}{\cdot}$ symbol shows the upper convected derivative~\cite{hinch2021oldroyd}. The yielding criteria is based on the deviatoric part of the polymeric stress tensor, defined as $\bar{\boldsymbol{\tau}}_d = \bar{\boldsymbol{\tau}}^p - (1/2)\text{tr}(\bar{\boldsymbol{\tau}}^p)I$, and on the following matrix norm $\lVert \boldsymbol{\tau} \rVert = \sqrt{\tau_{xx}^2 + \tau_{yy}^2 + 2\tau_{xy}^2}$. Note that this matrix norm was chosen as in the original paper by \citet{Saramito2009}, but other authors often choose a definition that introduces an additional factor of $\sqrt{2}$. Gravity is neglected in equation~\eqref{eq:momentum}, since the Bond number that characterizes our experiments is $\mathrm{Bo} = g\,\mathcal{L}^2\rho/\sigma \approx 0.1$ (for $g = 9.81\,m/s^2$), indicating that capillary effects remain dominant. Nevertheless, gravitational effects may already be non-negligible in this regime and will be examined in future work, particularly in light of previous studies that demonstrated a significant role of gravity in the spreading of yield-stress droplets (see e.g. \cite{DAngelo2022,heitmeier2025spreading}).

Some limits of equations \eqref{eq:momentum}-\eqref{eq:constitutive} are worth mentioning. In the limit $\mathrm{De} \rightarrow 0$, the elastic contribution vanishes and equation~\eqref{eq:constitutive} reduces to the steady viscoplastic form of a Herschel–Bulkley material. If $\mathcal{J} = 0$, we obtain the Oldroyd-B-like model with the addition of a shear-thinning element coming from the power index $n$. When $\mathcal{J} \rightarrow \infty$, the plastic element is never triggered and the material behaves like a Kelvin-Voigt elastic solid.

As described in our previous work \citep{Franca2024}, the system of equations~\eqref{eq:momentum}–\eqref{eq:constitutive} is solved using the open source code Basilisk C, which employs a finite-volume discretization on an adaptive quadtree grid, coupled with a volume-of-fluid (VOF) interface-capturing method. In the VOF method, the interface is tracked implicitly by a tracer field $c(\bar{\boldsymbol{x}}, t)$ which equals 1 inside the filaments, 0 in the outer fluid and values between 0 and 1 at the interface. Through this tracer field, the viscosity and density functions in equation \eqref{eq:momentum} are
\begin{equation}
\bar{\rho}  = c + (1 - c)\frac{\rho_a}{\rho}, \qquad
\bar{\eta}  = c + (1 - c)\frac{\eta_a}{\eta_s},
\label{eq:rho_mu}
\end{equation}
where $\rho_a$ and $\eta_a$ are, respectively, the density and viscosity of the ambient material around the filaments. In all simulations, we use the ratios ${\rho_a}/{\rho} = {\eta_a}/{\eta_s} = 10^{-2}$, which keeps the influence of the outer medium minimal.

Under the thin layer of fluid, a no-slip boundary condition is imposed at the bottom of the domain $(z = 0)$, and a symmetry boundary condition is applied at the left boundary $(x = 0)$. Consequently, only one filament needs to be simulated. The computational domain is a nondimensional box of height and width 5, and adaptive meshing is performed with a maximum level of 9, resulting in cells of size $\bar{\Delta}_x = 5/2^9 \approx 0.01$ (100 cells within the length scale of the problem). 

The initial shape of a single filament is given by the nondimensional height function
\begin{equation}
\bar{h}(\bar{x}) = \frac{35}{64}\frac{1}{\bar{R}_0} \cdot \max{\left[ 0, 1 - \left(\frac{\bar{x} - \bar{R}_0}{\bar{R}_0}\right)^2 \right] }^3 + \bar{h}_\infty,
\label{eq:initial_shape}
\end{equation}
where $\bar{R}_0$ is the radius of the filament base. In the main body of this paper, we always choose a filament with unity radius $\bar{R}_0 = 1$. Finally, the prefactor $35/64$ is calculated to guarantee that the combined area of both filaments is $\mathcal{L}^2 = 1$, as necessary by our choice of length scale in equation~\eqref{eq:vel_time_scale}.

\subsection{Scaling argument for the final bridge height}
\label{sec:theory}

% \color{blue}
% {
If the fluid layer yields fully during its evolution, we expect that, at late times, it will relax to a marginally-yielded state in which capillary forces are balanced by the resistive force due to the yield stress. In this case, we can predict how the final bridge height scales with the yield stress and surface tension by appealing to an order-of-magnitude force balance on a marginally-yielded layer. 

For this scaling argument, we consider the case in which the final layer depth has a local minimum at the bridge. Moreover, we assume that the final location of the maximum layer depth is close to $x=R_0$, the initial location of the center of the filament; we find that this is a good approximation for a number of simulations with viscoplastic fluid layers (e.g., Figure \ref{fig:vp_dynamics}B below), although it may not hold if $\mathcal{J}$ is very close to zero. The final average depth of the filament in $0\leq x\leq R_0$ is then given by
\begin{equation}
	\mathcal{H}_{\mathrm{avg}} \approx \frac{\mathcal{L}^2}{4R_0}.
\label{eq:h_avg_sclaing}
\end{equation}

The size of the driving capillary force on the fluid in $0\leq x\leq R_0$ can be estimated as 
\begin{equation}
	F_{\sigma}\sim \sigma\kappa \mathcal{H}_{\mathrm{avg}},\label{Fsigscl}
\end{equation}
where $\kappa$ is the free-surface curvature around the bridge. We estimate the curvature to be of order
\begin{equation}
	\kappa \sim \frac{\mathcal{H}_{\mathrm{avg}}-\mathcal{H}_{\mathrm{bridge}}}{R_0^2}.\label{kappascl}
\end{equation}
The resistive force due to the yield stress, $F_y$, which opposes the driving capillary force, is of order $F_y\sim\tau_yR_0$. Balancing $F_\sigma$ and $F_y$, while using \eqref{Fsigscl} and \eqref{kappascl}, gives
\begin{equation}
	\sigma \mathcal{H}_{\mathrm{avg}}(\mathcal{H}_{\mathrm{avg}}-\mathcal{H}_{\mathrm{bridge}}) \sim \tau_yR_0^3,
\end{equation}
which suggests
\begin{equation}
	\frac{\mathcal{H}_{\mathrm{bridge}}}{\mathcal{L}} \approx \Omega_1\frac{\mathcal{L}}{R_0}-\Omega_2\mathcal{J}\left(\frac{R_0}{\mathcal{L}}\right)^4,
	\label{eq:theory_scaling_1}
\end{equation}
for some prefactors, $\Omega_1$ and $\Omega_2$, that do not depend on $\mathcal{J}$. Thus, \eqref{eq:theory_scaling_1} indicates a linear decrease of the final bridge height with the plastocapillary number. 

We can estimate $\Omega_1$ by noting that if the layer's peak is near to $x=R_0$ for relatively small values of $\mathcal{J}$, then we expect $\mathcal{H}_{\mathrm{bridge}}\approx \mathcal{H}_{\mathrm{avg}}$ when $\mathcal{J}$ is small, suggesting $\Omega_1\approx\tfrac{1}{4}$ after using $\mathcal{H}_{\mathrm{avg}}$ from eq.~\eqref{eq:h_avg_sclaing}. We will compare \eqref{eq:theory_scaling_1}, using $\Omega_1=\tfrac{1}{4}$, with results from numerical simulations in the following section. The second prefactor, $\Omega_2$, %which may depend on $x_c$ and $\mathcal{L}$, 
is not determined via this scaling argument. 
%The scaling \eqref{eq:theory_scaling_1} reflects the fact that there is some dependence on the final bridge height depends of the area and initial separation of the filaments; we explore the impact of the initial conditions on the layer's evolution in detail in appendix \ref{sec:shape_dependence}. 
We expect \eqref{eq:theory_scaling_1} to provide a good prediction for the dependence of $\mathcal{H}_{\mathrm{bridge}}$ on $\mathcal{J}$ for purely viscoplastic layers, \emph{i.e.}, in the limit $\mathrm{De}\rightarrow0$, except when $\mathcal{J}$ is so large that the whole layer does not yield, or when $\mathcal{J}$ is so small that the peak layer height does not remain close to $x=R_0$. For larger $\mathrm{De}$, significant sub-yield elastic deformation is possible, so a simple balance between capillary stresses and yield stress is no longer likely to adequately describe the layer's late-time configuration. 
% }

\color{black}

\section{Results}

\subsection{Viscoplastic Limit: $\mathrm{De} \rightarrow 0$}
\label{sec:purely_vp}

\begin{figure}[t!]
\centering
\includegraphics[width=1\textwidth]{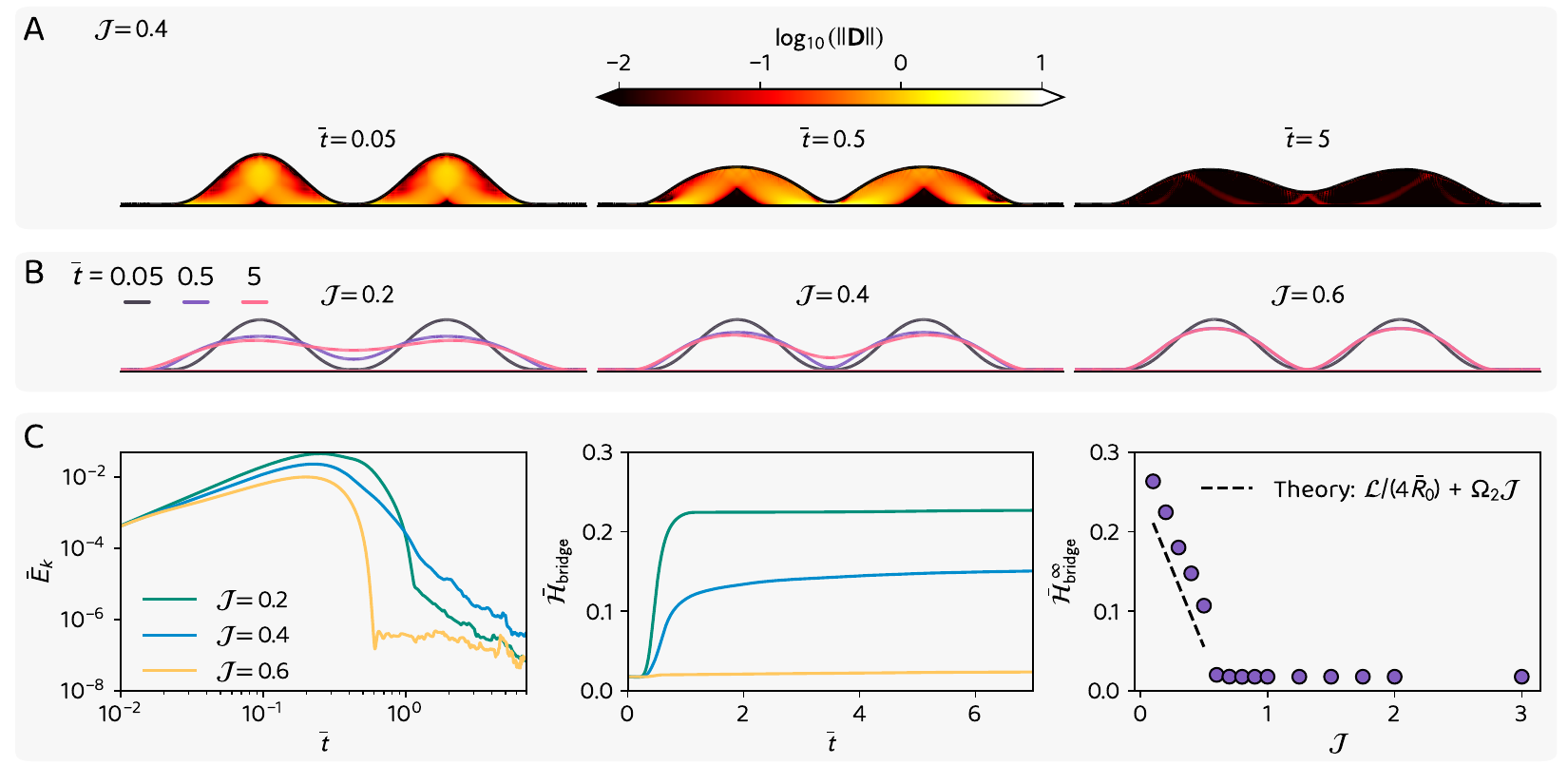}
\caption{Merging dynamics of quasi-viscoplastic filaments ($\mathrm{De} = 10^{-3}$). Panel A shows a time evolution of the shear rate for filaments with $\mathcal{J} = 0.4$: large plug regions form over time and coalescence stops eventually. Panel B illustrates the time evolution of filaments for three values of $\mathcal{J}$, illustrating the pronounced arrested coalescence for large $\mathcal{J}$. Panel C contains the time evolution of kinetic energy and the bridge height, showing how the bridge reaches an arrested shape at late-time in the presence of yield-stress. Panel C also shows the final bridge height as a function of $\mathcal{J}$, compared with the linear scaling predicted in equation~\eqref{eq:theory_scaling_1}. The prefactor $\Omega_2 = -0.4$ was obtained by fitting the slope of the numerical data. In all simulations we keep fixed $\mathrm{Oh}_s = 1/90$ and $\mathrm{Oh}_p = 8/90$.}
\label{fig:vp_dynamics}
\end{figure}

We begin with a limit in which analytical results are more accessible, when $\mathrm{De}\rightarrow0$ (see above). Figure~\ref{fig:vp_dynamics} illustrates the coalescence dynamics of two neighboring filaments in this viscoplastic limit. In this nearly purely plastocapillary regime~\cite{Jalaal2021,Franca2024}, the material presents minimal elasticity and behaves almost as a Herschel--Bulkley fluid, where the interplay between capillary and yield stresses governs the overall dynamics. As such, the relevant control parameter in this regime is the plastocapillary number $\mathcal{J}$. Note that, using $\mathrm{De} = 0$ would pose a well-known numerical challenge associated with pure viscoplasticity, requiring regularized models such as the one proposed by \citet{Papanastasiou1987}. In this work, we opted to regularize the viscoplastic limit by adding a very small amount of elasticity instead $(\mathrm{De} = 10^{-3})$. Hence, we use the same framework for our viscoplastic and elasto-viscoplastic simulations (to follow). 
% which makes the transition to full EVP simulations in the next section smoother, since only a single model is used for both the VP and EVP sections. 
See appendix \ref{section:regularization} for additional comments on the usage of small elasticity as a method of numerical regularization.

Panel A of Figure~\ref{fig:vp_dynamics} shows the temporal evolution of the coalescence process for a representative case with $\mathcal{J} = 0.4$. The color field represents the magnitude of the deformation-rate tensor $ \textbf{D} = 1/2\ \left[ \nabla \textbf{u} + (\nabla\textbf{u})^T \right]$, with high values indicating the regions where the material is actively flowing and low values highlighting plug regions. Initially, at short times ($\bar{t} = 0.05$), the two filaments are separate and deformation is initiated at the contact lines between the filaments and the underlying film, where capillary pressure is higher due to higher curvatures. As the bridge forms and grows ($\bar{t} = 0.5$), a strong shear zone allows flow in the narrow neck region connecting the filaments, while a portion of the filament's bulk remains unyielded. At later times $(\bar{t} = 5)$, the deformation progressively decays and the entire fluid becomes yielded, with a bridge height that saturates to a finite value. In this scenario the final configuration is reached when a balance between the capillary forces and the resisting force from yield stress is obtained.

Panel B compares the interface evolution for three plastocapillary numbers, $\mathcal{J} = 0.2$, $0.4$, and $0.6$. The contours show the interface position at several nondimensional times. For low yield stress ($\mathcal{J} = 0.2$), surface tension dominates and the bridge grows substantially, nearly merging the two filaments into a single one. As $\mathcal{J}$ increases, the material’s resistance to flow becomes more pronounced: the deformation is confined to a smaller region near the bridge, and the final shape retains a visible separation between the filaments. For $\mathcal{J} = 0.6$, the capillary stresses are unable to fully overcome the yield stress, and only a thin, arrested bridge forms. These results demonstrate how increasing the plastocapillary number progressively suppresses the extent of coalescence, transitioning the response from fluid-like to solid-like.

To further quantify the effect of plastocapillary number on the final bridge height, we look at the temporal variation of this value and the kinetic energy of this regularized system to ensure we arrived at an asymptotically large time. In panel C, the kinetic energy of the two filaments is plotted over time.
% to illustrate that these simulations are converging to a stop before full coalescence is reached. 
Due to numerical approximations, the simulation never reaches exactly zero kinetic energy. With this limitation in mind, we choose to consider that the coalescence has fully stopped when the nondimensional kinetic energy $\bar{E}_k = E_k / ( \sigma \mathcal{L}^2 )$ equals $10^{-6}$, with $E_k = \int_\Omega (1/2)\rho\lVert\mathbf{u}\rVert^2 d\Omega$ integrated over the area of the filament cross-section. In panel C, we also show the evolution of the bridge height during the coalescence process. As the kinetic energy rapidly decreases, we see that the bridge height also clearly stops growing before full coalescence has happened. For higher values of $\mathcal{J}$ the initial increase in kinetic energy is lower, such that it reaches the stoppage criteria quickly before any significant coalescence has happened. 

Finally, we also visualize in panel C the final bridge height $\mathcal{H}_{\text{bridge}}^{\infty}$ as a function of $\mathcal{J}$ for this viscoplastic limit scenario. %For $\mathcal{J\rightarrow} \ 0$, the bridge height tends to the peak of a single fully coalesced Newtonian droplet. 
As $\mathcal{J}$ is increased from $0.1$ to $0.6$,  the bridge height decreases linearly, as predicted by the scaling argument \eqref{eq:theory_scaling_1}. This linearity seems to be violated near the critical $\mathcal{J}$. At this critical value of $\mathcal{J}$ the bridge height plateaus at a constant value, at which the capillary pressure near the bridge from the initial condition is not strong enough to overcome the yield-stress and generate flow. This critical $\mathcal{J}$ is evidently dependent on some features of the initial geometry, including the location of the filament center and its base radius, $\bar{R}_0$, as suggested by \eqref{eq:theory_scaling_1}. In fact, the impact of the initial shape is rather significant and may be of great interest of applications like 3D printing. We further investigated these effects in appendix~\ref{sec:shape_dependence}.

\subsection{Effect of Elasticity}
\label{sec:effect_elasticity}

We now investigate how the addition of elasticity changes the picture previously shown in Section~\ref{sec:purely_vp}. This is done by increasing the Deborah number, which introduces a nondimensional timescale associated with polymeric stress responses. This timescale characterizes how long it takes for polymeric stress to develop in the material when subjected to deformation, and how long it takes for these stresses to relax.

\begin{figure}[htbp]
\centering
\includegraphics[width=1\textwidth]{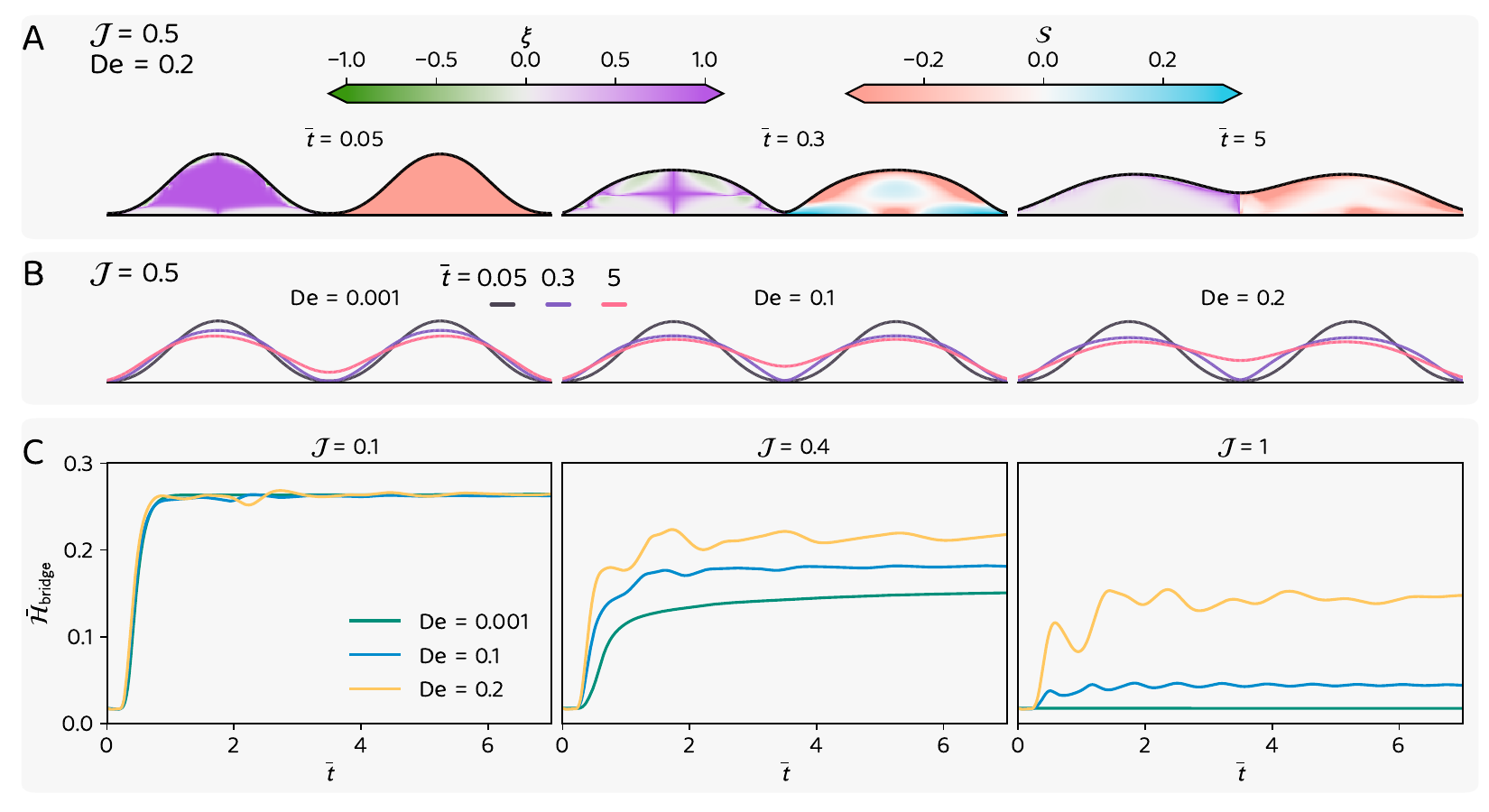}
\caption{Merging dynamics of elastoviscoplastic filaments. Panel A shows time evolution of the flow type $(\xi)$ and solidification $(\mathcal{S})$ parameters for filaments with $\mathcal{J} = 0.5$ and $\mathrm{De} = 0.2$. We note that specific flow regimes can be overlapped with yielded and unyielded areas. Panel B illustrates the interface evolution of filaments for three values of $\mathrm{De}$ and fixed $\J = 0.5$, illustrating that going from a rigid to an elastic solid allows for additional deformation and stronger coalescence. Panel C presents the bridge height over time for various $\mathrm{De}$ and $\J$, showing that elasticity plays a greater role in the solid regime than in fluidized one. In all simulations we keep fixed $\mathrm{Oh}_s = 1/90$ and $\mathrm{Oh}_p = 8/90$ fixed.} 
\label{fig:evp_dynamics}
\end{figure}

In panel A of Figure \ref{fig:evp_dynamics} we show the snapshots of a single simulation with $\J = 0.5$ and $\mathrm{De} = 0.2$. We visualize the internal flow in the filaments by plotting the so-called flow-type parameter $\xi = \frac{|\textbf{D}| - |\mathbf{\Omega}|}{|\textbf{D}| + |\mathbf{\Omega}|}$, where $\textbf{D}$ and $\mathbf{\Omega}$ are the deformation rate and vorticity tensors (the symmetric and antisymmetric components of $\nabla \mathbf{u}$), respectively: $ \textbf{D} = 1/2\ \left[ \nabla \textbf{u} + (\nabla\textbf{u})^T \right]$ and $\mathbf{\Omega} = 1/2\left[ \nabla \textbf{u} - (\nabla\textbf{u})^T \right]$. Initially $(\bar{t} = 0.05)$, we see areas with $\xi = 0$ near the edges of the filaments and the solid surface, indicating strong shear flow induced by high curvature near the bridge area and the contact line between filament and thin-film. Over time $(\bar{t} = 0.3)$, this shear zone grows as more of the filaments deform and the bridge grows. At this stage, only the center of the filaments present a localized area with $\xi = 1$, indicating extensional flow. At later time the flow beneath the bridge also shows the signature of extensional flows but eventually, the whole process stops due to yield-stress and a final bridge height is obtained, similarly to purely viscoplastic scenario. In each snapshot we also visualize the magnitude of the polymeric stresses through the scalar $\mathcal{S} = \log(\norm{\bm{\bar{\tau}}_{d}}) - \log{(\J)}$, with positive values indicating (viscoelastic) fluid regions $(\norm{\bm{\bar{\tau}}_{d}} > \J)$ and negative values (viscoelastic) solid regions. Overall, the process of coalescence leads to a highly heterogeneous field where solid-like and liquid-like behavior co-exists. The regions of strong shear near the bridge and the contact lines clearly yields the flow and in the center of the droplet the strong extension, also, leads to a yielded region. During the deformation, close to the surface of the filaments, a localized solid areas develops as these regions locally experience a solid-body rotation with small polymeric stresses. As the coalescence slows down, the polymeric stresses in the whole filament decrease and the entire material becomes a solid.

Panel B of figure \ref{fig:evp_dynamics} shows snapshots of the interface for three different simulations with, respectively, $\mathrm{De} = 0.001, \ 0.1, \ 0.2$. The plastocapillary number is kept fixed at $\J = 0.5$ in all three cases of this panel. Overall, a similar behavior is observed as in the purely VP scenario: the filaments start to coalesce and the bridge grows to a final shape. We note, however, that increasing the Deborah number at fixed plastocapillary leads to larger arrested bridge heights. This occurs because finite elasticity in the Saramito model allows the material to deform elastically before yielding, enabling additional interface motion even when the local polymeric stress is still below the yield threshold. As the bridge forms, capillary stresses generate elastic tension that relaxes only on time scales of order $\mathcal{O}(\mathrm{De})$, and therefore continue to drive coalescence after the curvature has decreased to values for which a purely viscoplastic material would already arrest. The final equilibrium condition involves a balance between capillary, elastic, and yield stresses; the presence of an elastic contribution reduces the curvature required for arrest, resulting in a smoother interface and a larger final bridge height.

An interesting dynamical feature of the EVP solution (hard to visualize in the selected snapshots of panel B) is the the filaments oscillation before reaching their final shape. This is better shown in panel C, where we plot the bridge height over time for various $\J$ and $\mathrm{De}$, and also on supplementary video I. When yield stress ($\J$) is small, the material mainly behaves like a Newtonian fluid for small $\mathrm{De}$ and a viscoelastic liquid for larger values of $\mathrm{De}$. In both limits, the material experiences significant amount of deformation and the yield stress only plays a role in long times to solidify the filaments. Hence, increasing elasticity does not affect the final bridge height. While the final shape is the same, we note for $\mathrm{De} = 0.2$ that oscillations are present at finite time, eventually dying out as energy is dissipated through viscosity. For higher yield-stress ($\J = 0.4$ and $\J = 1$), we note that increasing the Deborah number not only affects dynamics (through oscillations) but the final shape is also changed. At high yield-stress, by increasing $\mathrm{De}$ we transition from a viscoplastic Herschel-Bulkley material to a Kelvin-Voigt elastic solid. The VP material experiences little to no deformation, since it is a rigid solid at small stresses. The Kelvin-Voigt solid, however, can elastically deform even at small stress, leading to an increase in bridge height. Therefore, the effect of $\mathrm{De}$ is pronounced for high values of $\J$ due to this transition from a fully rigid solid to an elastic solid.
In the section below, we focus on this viscoelastic solid regime and in particular look into the oscillations observed in the simulations. 

\subsection{Kelvin-Voigt limit: elastic oscillations at $\mathcal{J \rightarrow \infty}$}
\label{sec:kv_oscillations}

In the limit of very large plastocapillary number ($J \rightarrow \infty$), the Saramito model reduces to a Kelvin--Voigt viscoelastic solid. 
% When yielding is effectively suppressed, the evolution of two coalescing filaments is governed by the competition between elastic stresses, which store and release energy, and viscous dissipation, which damps motion. 
Our simulations reveal that, in this limit, the bridge height does not relax monotonically toward its arrested configuration. Instead, after a short geometric adjustment, the system exhibits exponentially damped oscillations around its final shape. This behaviour is characteristic of Kelvin--Voigt solids undergoing shape relaxation under capillary driving forces.

To rationalize these oscillations, we derive a minimal model in which perturbations of the filament height are represented as the displacement of a damped harmonic oscillator. Note that for this particular purpose, we focus on the droplet height in the centre, rather than the bridge height, since in the computations it shows a more pronounced oscillation for a larger range of $\mathrm{De}$. This can be better visualized in supplementary videos II and III. For sufficiently small Deborah numbers, the timescale of elastic relaxation is significantly smaller than capillarity, such that the restoring force in the system can be considered to come purely from elastic stresses. For infinite $\mathcal{J}$, Saramito's constitutive equation~\eqref{eq:constitutive} reduces to
$
\overset{\nabla}{\bar{\boldsymbol{\tau}}^p} = \left( \mathrm{Oh}_p/\mathrm{De} \right) \, \bar{\dot{\boldsymbol{\gamma}}},
$
which implies a purely elastic stress with nondimensional elastic modulus $\mathrm{Oh}_p/\mathrm{De}$. The viscosity in the system will come entirely from the solvent stress in equation \eqref{eq:momentum}, given by $\tau_s = \mathrm{Oh}_s\bar{\dot{\boldsymbol{\gamma}}}$. As such, we model the droplet height as a damped harmonic oscillator expressed by the equation
\begin{equation}
\ddot{\bar{\mathcal{H}}}_{\mathrm{peak}} + \mathrm{Oh}_s\dot{\bar{\mathcal{H}}}_{\mathrm{peak}} + \frac{\mathrm{Oh}_p}{\mathrm{De}}\bar{\mathcal{H}}_{\mathrm{peak}} = 0,
\label{eq:kv_harmonic_oscillator}
\end{equation}
where $\bar{\mathcal{H}}_{\mathrm{peak}}$ is the height of a filament.

Equation \eqref{eq:kv_harmonic_oscillator} will admit solutions of the form
\begin{equation}
\bar{\mathcal{H}}_{\mathrm{peak}}(\bar{t}) = Ae^{-\lambda  \bar{t}}\sin{\left( \bar{\omega}  \bar{t} + \phi\right) } + B,
\label{eq:solution_oscillation}
\end{equation}
where
\begin{equation}
\lambda =  \mathrm{Oh}_s, \qquad \bar{\omega} = \left(\frac{\mathrm{Oh}_p}{\mathrm{De}}\right)^{1/2}\left(1 - \frac{\mathrm{Oh}_s^2}{4}\frac{\mathrm{De}}{\mathrm{Oh}_p}\right)^{1/2} \approx \left(\frac{\mathrm{Oh}_p}{\mathrm{De}}\right)^{1/2},
\label{eq:sol_kv_oscillator}
\end{equation}
and $A$ and $\phi$ are dependent on the initial conditions, while $B$ determines the final height after full coalescence. The approximation used for $\bar{\omega}$ in equation~\eqref{eq:sol_kv_oscillator} is reasonable whenever an underdamped oscillation is present, that is, $(\mathrm{Oh}_s^2 \ \mathrm{De})/(4 \ \mathrm{Oh}_p) << 1$.

From solution \eqref{eq:sol_kv_oscillator}, we conclude that the model predicts a natural oscillation frequency approximately proportional to $(\mathrm{Oh}_p/\mathrm{De})^{1/2}$ and a decay rate that only has a linear dependence on the solvent Ohnesorge number.

\begin{figure}[htbp]
\centering
\includegraphics[width=1\textwidth]{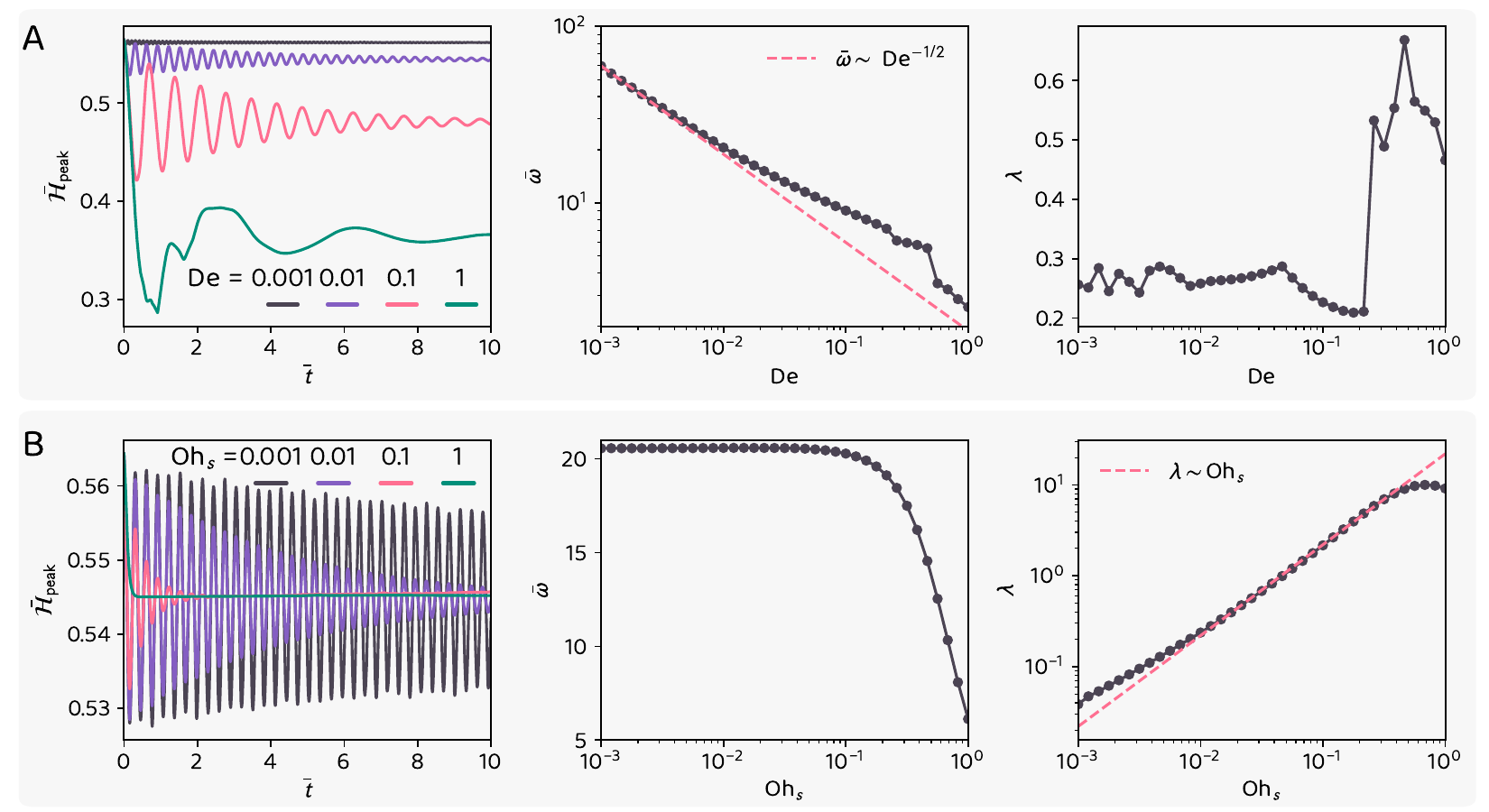}
\caption{Oscillatory dynamics of Kelvin-Voigt filaments during coalescence. We use $\mathcal{J}=10^4$ in all simulations, guaranteeing that the material is always unyielded. Panel A contains a sweep of $\mathrm{De}$ for fixed $\mathrm{Oh}_s = 1/90$ and $\mathrm{Oh}_p = 8/90$, illustrating the harmonic oscillator behaviour with a frequency determined by $\mathrm{De}$. Panel B presents a sweep of $\mathrm{Oh}_s$ for fixed $\mathrm{De} = 0.01$ and $\mathrm{Oh}_p = 8/90$, showing that the oscillator decay depends on $\mathrm{Oh}_s$, but the frequency does not. The filament shapes during these oscillations can be seen in supplementary videos II and III.}
\label{fig:kv_oscillations}
\end{figure}

Figure~\ref{fig:kv_oscillations} summarizes the oscillatory dynamics observed in the Kelvin--Voigt limit within our simulations. 
Panel~A explores the effect of elasticity by varying the Deborah number while holding the solvent 
viscosity fixed $\mathrm{Oh}_s = 1/90$. The first panel shows the temporal evolution of the droplet height for four 
representative values of $\mathrm{De}$ at multiple orders of magnitude. At small $\mathrm{De}$ the dynamics closely follow a simple 
exponentially damped sinusoid, consistent with our underdamped Kelvin--Voigt prediction \eqref{eq:solution_oscillation}. As 
$\mathrm{De}$ increases, however, the waveform progressively departs from a pure sinusoid: the 
oscillations become asymmetric and the instantaneous frequency exhibits a noticeable drift, indicating 
that the effective restoring force is no longer dominated by the linear elastic contribution alone. For each simulation, we fit the droplet height over time to the solution \eqref{eq:solution_oscillation} in order to obtain $\bar{\omega}$ and $\lambda$. The middle panel shows the obtained oscillation period $\bar{\omega}$, verifying that it scales as 
$\mathrm{De}^{-1/2}$ for $\mathrm{De}\ll 1$, in agreement with the scaling obtained from the 
damped harmonic oscillator model. For larger $\mathrm{De}$ this trend breaks down as capillary and 
geometric nonlinearities begin to contribute comparably to the restoring force. Finally, the decay 
rate $\lambda$ (right panel) is found to be nearly constant for small $\mathrm{De}$, as expected when 
viscous dissipation is controlled purely by the solvent dashpot ($\mathrm{Oh}_s$). 

Panel~B isolates the effect of viscous dissipation by fixing $\mathrm{De}=10^{-2}$ and varying the 
solvent Ohnesorge number. The left panel shows the corresponding height dynamics, where increasing 
Oh$_s$ leads to progressively stronger damping and a faster decay of the oscillations. Consistent with 
the Kelvin--Voigt prediction, the oscillation period (middle panel) remains essentially independent of 
viscosity for $\mathrm{Oh}_s\ll 1$, illustrating that the natural frequency is set primarily by the elastic 
modulus. The decay rate $\lambda$, plotted in the right panel, grows 
approximately linearly with Oh$_s$ in the small-viscosity regime, also as expected from the 
Kelvin--Voigt dashpot contribution. At larger Oh$_s$, the decay departs from linearity as the motion 
transitions toward an overdamped regime in which oscillations are strongly suppressed.

\subsection{Experiments, Theory and EVP Computations}

\begin{figure}[t!]
\centering
\includegraphics[width=1\textwidth]{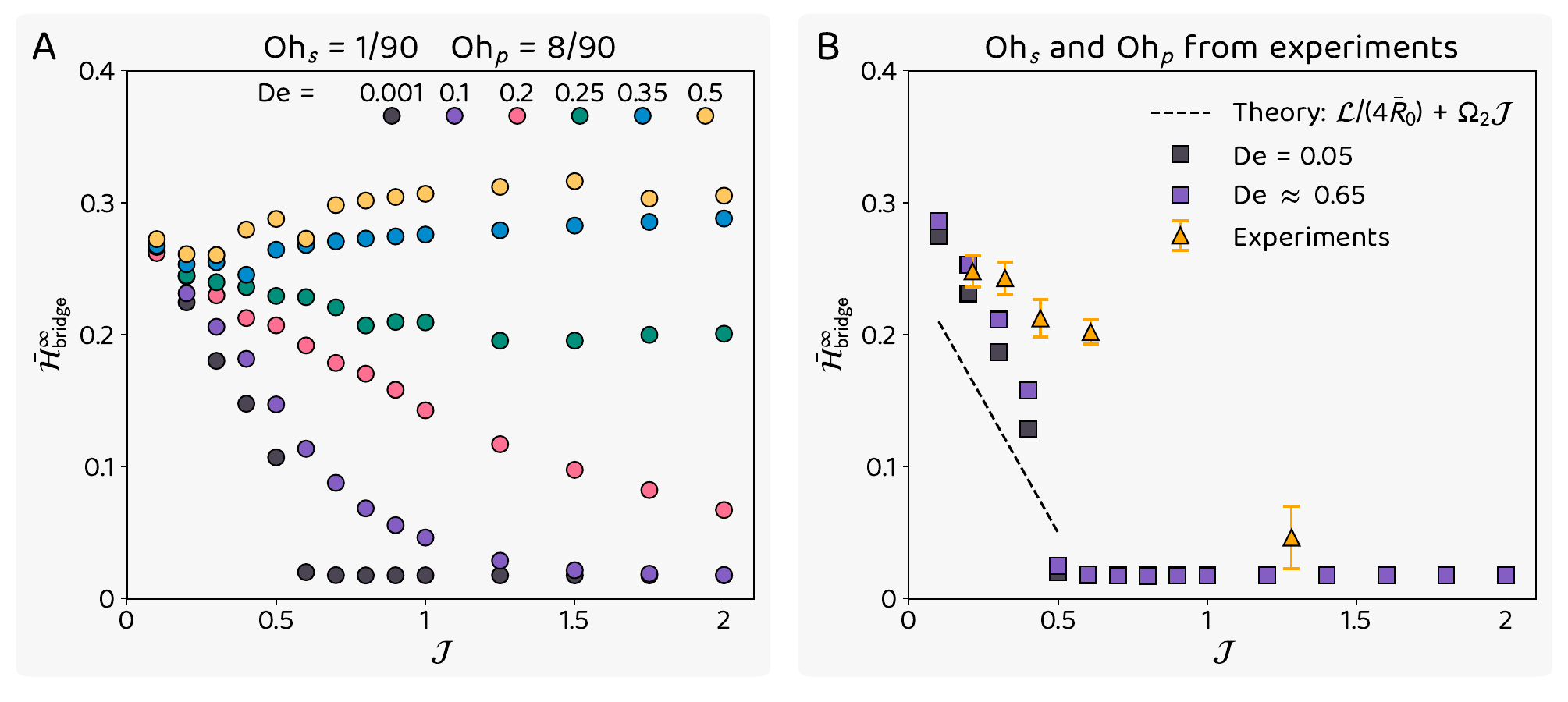}
\caption{Steady-state bridge height as a function of $\J$ and $\mathrm{De}$. Panel A shows results for small Ohnesorge numbers, where fast and inertial coalescence takes place. These results illustrate that elasticity is more influential to the steady-state solution when $\J$ is high. Panel B shows results from Carbopol experiments and simulations with comparable (high) Ohnesorge numbers. The elasticity of Carbopol has little influence on the final shape of filaments. The theory line is the same calculated for the purely viscoplastic section in figure \ref{fig:vp_dynamics}.}
\label{fig:evp_scaling}
\end{figure}

We bring together the different strands of our study and compare the final bridge height obtained from EVP simulations, purely viscoplastic (VP) theory, and experiments.
In panel A of figure \ref{fig:evp_scaling}, we present only numerical results obtained with low values of the Ohnesorge numbers (which are not compatible with experiments, see appendix~\ref{section:app:rheo}), such that inertial features may be present. For small values of $\mathrm{De}$, the filaments approach a purely viscoplastic spreading. As such, the scaling law in this regime tends to the one presented in figure \ref{fig:vp_dynamics}. 

For small values of $\mathcal{J}$, we observe that changes in $\mathrm{De}$ do not lead to significant impact on the final bridge height (or at least the trend). As commented previously, the transition from a Newtonian to a viscoelastic fluid makes little effect to the final bridge height, since both scenarios do not solidify and fully spread. 

On the other hand, for high values of $\mathcal{J}$, increasing $\mathrm{De}$ drives a transition from a rigid Herschel–Bulkley solid to a viscoelastic Kelvin–Voigt solid. For $\mathrm{De}=0$, Eq.~\eqref{eq:constitutive} yields a polymeric stress that is purely viscoplastic and, due to the large yield stress ($\mathcal{J}$), the filaments remain stuck at their initial positions and no coalescence occurs. For large values of $\mathrm{De}$, the timescale over which polymeric stresses develop is increased. Before this timescale is reached, the polymeric stresses remain small, so that most (or all) of the filaments are unyielded and behave locally as a Kelvin–Voigt solid. These Kelvin–Voigt regions experience dissipation only through the solvent viscosity $\mathrm{Oh}_s$, allowing them to coalesce faster than their fluidized counterparts, which additionally dissipate through the polymeric viscosity $\mathrm{Oh}_p$. Coalescence then proceeds and eventually arrests due to a balance between elastic and capillary stresses.

In panel B of figure \ref{fig:evp_scaling}, we show simulations and experiments with high Ohnesorge numbers, comparable to those in our experiments. We note that in panel B of figure \ref{fig:evp_scaling}, the value for the Ohnesorge number is not a constant between all points. As shown in table \ref{table:rheology_parameters}, the Deborah number and the shear-thinning index $n$ are similar in all 5 Carbopol samples, and approximately $\mathrm{De} \approx 0.65$ and $n \approx  0.5$. However, changes in Carbopol concentration are observed to significantly influence the values of $\mathrm{Oh}_p$, $\mathrm{Oh}_s$ and $\mathcal{J}$. Since experimental data were available only for a discrete set of 5 samples, the corresponding Ohnesorge numbers $\mathrm{Oh}_s$ and $\mathrm{Oh}_p$ used in each simulation were obtained by linear interpolation using the two experimental points with the closest values of $\mathcal{J}$. This procedure ensures that the simulated values of $\mathrm{Oh}_s$ and $\mathrm{Oh}_p$ vary with $\mathcal{J}$, consistently with this experimentally-observed dependence. Comparing the simulations from $\mathrm{De} = 0.05$ and $\mathrm{De} = 0.65$, we initially observe that the elasticity introduced by the larger $\mathrm{De}$ does not significantly alter the final shape of these filaments. In this high $\mathrm{Oh}$ scenario, the initial elastic oscillations observed in figure \ref{fig:evp_dynamics} are completely suppressed, and simply a very viscous and slow coalescence happens until the bridge stops growing. We also note that the scaling looks very similar to the small $\mathrm{De}$ limit in Panel A, which shows that, whenever elasticity is not significant, changing viscosity does not significantly alter the final shape, but mainly delays how long it will take to reach it. The experimental data also present the main observations from our simulations and theory: a maximum bridge height $\bar{\mathcal{H}}^\infty_{\max}$ attained at the lowest value of $\mathcal{J}$ tested, and a linear decrease of the bridge height with increasing $\mathcal{J}$. 
As discussed in Sec.~\ref{sec:theory}, the values of $\bar{\mathcal{H}}^\infty_{\max}$ and the critical $\mathcal{J}$ depend sensitively on the initial filament geometry. As a result, quantitative differences between experiments and simulations are expected, particularly at larger $\mathcal{J}$ where memory of the initial conditions is more pronounced. Nevertheless, the overall agreement at small $\mathcal{J}$ is good, where the influence of the initial shape is weaker and the trends observed in experiments and simulations closely align.
% As mentioned in section \ref{sec:theory}, the values for $\bar{\mathcal{H}}^\infty_\text{max}$ and for the critical $\mathcal{J}$ are strongly dependent on the initial shape of the filaments, such that they do not match between our experiments and simulations. 

% To ask Hugo:\\
% - is it really stopped?\\
% - shape effects? as the linearity breaks at high J?\\
% - let's add a couple of data points of Daniel - I forgot how did it look like whn you did this last time - I uploaded the final theses here too\\
% - early morning crazy idea: of we add just a tiny bit of elastiity (De) we kind of regulirize pure plasticity buuut- the flow actually stops, no? so maybe can test that also and compar with lubrication analysis.\\
% - late-night crazy idea: define the stoppage criteria only for the area close to the bridge, e.g., on the bridge velocity.

\section{Conclusion \& Outlook}
\label{sec:conclusion}

We have studied the coalescence of two neighbouring filaments of a yield-stress fluid on a solid surface, a process that can be viewed as a two-dimensional analogue of sessile-droplet coalescence in which the merger arrests at a non-trivial final shape. Motivated by multilayer direct-ink-writing (DIW) and related additive-manufacturing settings, we combined a scaling theory for the arrested bridge height in the viscoplastic limit, elasto-viscoplastic (EVP) direct numerical simulations based on the Saramito constituve model, and experiments, featuring an OCT-based surface profilometry on Carbopol filaments. Our focus was on the late-time final bridge height and the plastocapillary regimes under which capillarity can overcome plastic resistance.

In the quasi-viscoplastic regime, the dynamics evolve towards a marginally yielded configuration where capillary driving is balanced primarily by yield-stress resistance, leading to a final bridge height that decreases approximately linearly with the plastocapillary number over an extended range of parameters. This picture naturally breaks down near a critical plastocapillary number, beyond which the initial capillary pressure is insufficient to trigger flow and the bridge essentially fails to grow. As we discussed in appendix~\ref{sec:shape_dependence}, this behaviour is not a universal material property but is strongly shaped by the effective initial curvature between filaments, an observation that is particularly relevant for DIW, where deposition history sets the initial condition. When elasticity is introduced, the outcome depends strongly on whether the material is effectively fluid-like or solid-like: for small yield stress (small $\mathcal{J}$), elasticity primarily modifies transients (including oscillations) while leaving the final bridge height largely unchanged, whereas for larger $\mathcal{J}$ elasticity can materially increase the arrested bridge height by enabling sub-yield deformation and shifting the late-time balance towards an elasto-capillary–plastic competition. These trends connect naturally to recent work showing that elasticity can influence coalescence more broadly~\cite{Varma2022,Dekker2022} and to emerging studies of viscoplastic sessile coalescence~\cite{Kern2022}, while placing the emphasis on filament geometries that are directly relevant to printing.

Overall, the agreement between experiments, EVP simulations, and the viscoplastic scaling is encouraging at low $\mathcal{J}$, where the memory of the initial condition is weaker and the dominant trends align well. At higher $\mathcal{J}$ the match becomes less quantitative, which is not surprising given the strong sensitivity of $\bar{\mathcal{H}}^\infty_{\max}$ and the apparent critical $\mathcal{J}$ to the initial filament shape, and the fact that, in printing, the initial geometry is itself a rheology–process outcome (nozzle height, traverse speed, extrusion rate, start/stop protocols, and the flow conditioning of the material inside the nozzle). Making this connection predictive, \emph{i.e.} linking process parameters to a well-posed initial condition for coalescence, remains an important next step for translating non-Newtonian filament coalescence physics into robust print guidelines.

Several extensions are natural. Many practical DIW workflows involve printing into or within a supporting yield-stress bath (embedded printing~\cite{grosskopf2018viscoplastic,friedrich2024simulated}). Extending our framework to an embedding medium with its own yield stress and elasticity is conceptually straightforward within the present two-phase formulation and would directly broaden applicability to embedded geometries. On the experimental side, OCT offers considerably more than static surface profiles: high-speed OCT could resolve the full time evolution of bridge growth, quantify transient ridge formation, and even flow visualisation or local microrheology~\cite{Jalaal2018}, which we did not exploit here. On the rheology side, incorporating and systematically probing extensional rheology in experiments would help clarify its role during early-time neck formation and subsequent relaxation~\cite{saengow2025stretching}. Additionally, on the modelling side, while the Saramito EVP model provides a convenient and well-tested route to couple yielding and elasticity, other constitutive descriptions~\cite{larson2015constitutive,saramito2021new,kamani2024brittle,pagani2024no,divoux2024ductile} may be needed for quantitative comparisons in the solid-like regime, including improved descriptions of sub-yield viscoelasticity, multiple relaxation modes, and (where relevant) thixotropy/aging. Exploring such refinements may provide a more complete predictive tool for deposition profiles and for mitigating residual ridges/grooves in printed surfaces.

% \section*{Acknowledgements}
% We are grateful to ...

\section*{Acknowledgement}
M.J. acknowledges support from the Vidi project Living Levers with file No. 21239, financed by the Dutch Research Council (NWO).
C. Oishi would like to thank the financial support given by the São Paulo Research Foundation (FAPESP) process numbers  2013/07375-0, 2023/14427-8 and  2024/04769-1 and the National Council for Scientific and Technological Development (CNPq), grants 305383/2019-1 and 307228/2023-1.
\printbibliography

@article{Rostami2025,
  title = {Coalescence of viscoelastic drops on a solid substrate},
  volume = {10},
  ISSN = {2469-990X},
  url = {http://dx.doi.org/10.1103/PhysRevFluids.10.063603},
  DOI = {10.1103/physrevfluids.10.063603},
  number = {6},
  journal = {Physical Review Fluids},
  publisher = {American Physical Society (APS)},
  author = {Rostami,  Peyman and Erb,  Alexander and Azizmalayeri,  Reza and Steinmann,  Johanna and Stark,  Robert W. and Auernhammer,  G\"{u}nter K.},
  year = {2025},
  month = jun 
}

@article{Papanastasiou1987,
  title = {Flows of Materials with Yield},
  volume = {31},
  ISSN = {1520-8516},
  url = {http://dx.doi.org/10.1122/1.549926},
  DOI = {10.1122/1.549926},
  number = {5},
  journal = {Journal of Rheology},
  publisher = {Society of Rheology},
  author = {Papanastasiou,  Tasos C.},
  year = {1987},
  month = jul,
  pages = {385–404}
}

@article{Franca2024,
  title = {Elasto-viscoplastic spreading: From plastocapillarity to elastocapillarity},
  volume = {6},
  ISSN = {2643-1564},
  url = {http://dx.doi.org/10.1103/PhysRevResearch.6.013226},
  DOI = {10.1103/physrevresearch.6.013226},
  number = {1},
  journal = {Physical Review Research},
  publisher = {American Physical Society (APS)},
  author = {Fran\c{c}a,  Hugo L. and Jalaal,  Maziyar and Oishi,  Cassio M.},
  year = {2024},
  month = mar 
}

@article{Saramito2009,
  title = {A new elastoviscoplastic model based on the Herschel–Bulkley viscoplastic model},
  volume = {158},
  ISSN = {0377-0257},
  url = {http://dx.doi.org/10.1016/j.jnnfm.2008.12.001},
  DOI = {10.1016/j.jnnfm.2008.12.001},
  number = {1–3},
  journal = {Journal of Non-Newtonian Fluid Mechanics},
  publisher = {Elsevier BV},
  author = {Saramito,  Pierre},
  year = {2009},
  month = may,
  pages = {154–161}
}

@article{Sanjay2021,
  title = {Bursting bubble in a viscoplastic medium},
  volume = {922},
  ISSN = {1469-7645},
  url = {http://dx.doi.org/10.1017/jfm.2021.489},
  DOI = {10.1017/jfm.2021.489},
  journal = {Journal of Fluid Mechanics},
  publisher = {Cambridge University Press (CUP)},
  author = {Sanjay,  Vatsal and Lohse,  Detlef and Jalaal,  Maziyar},
  year = {2021},
  month = jul 
}

@article{Dekker2022,
  title = {When Elasticity Affects Drop Coalescence},
  volume = {128},
  ISSN = {1079-7114},
  url = {http://dx.doi.org/10.1103/PhysRevLett.128.028004},
  DOI = {10.1103/physrevlett.128.028004},
  number = {2},
  journal = {Physical Review Letters},
  publisher = {American Physical Society (APS)},
  author = {Dekker,  Pim J. and Hack,  Michiel A. and Tewes,  Walter and Datt,  Charu and Bouillant,  Ambre and Snoeijer,  Jacco H.},
  year = {2022},
  month = jan 
}

@article{Dahiya2016,
  title = {Arrested coalescence of viscoelastic droplets: polydisperse doublets},
  volume = {374},
  ISSN = {1471-2962},
  url = {http://dx.doi.org/10.1098/rsta.2015.0132},
  DOI = {10.1098/rsta.2015.0132},
  number = {2072},
  journal = {Philosophical Transactions of the Royal Society A: Mathematical,  Physical and Engineering Sciences},
  publisher = {The Royal Society},
  author = {Dahiya,  Prerna and Caggioni,  Marco and Spicer,  Patrick T.},
  year = {2016},
  month = jul,
  pages = {20150132}
}

@article{Pawar2012,
  title = {Arrested coalescence of viscoelastic droplets with internal microstructure},
  volume = {158},
  ISSN = {1364-5498},
  url = {http://dx.doi.org/10.1039/C2FD20029E},
  DOI = {10.1039/c2fd20029e},
  journal = {Faraday Discussions},
  publisher = {Royal Society of Chemistry (RSC)},
  author = {Pawar,  Amar B. and Caggioni,  Marco and Hartel,  Richard W. and Spicer,  Patrick T.},
  year = {2012},
  pages = {341}
}

@article{Kaneelil2026,
  title = {Coalescence of viscoelastic sessile drops: the small and large contact angle limits},
  volume = {1026},
  ISSN = {1469-7645},
  url = {http://dx.doi.org/10.1017/jfm.2025.10878},
  DOI = {10.1017/jfm.2025.10878},
  journal = {Journal of Fluid Mechanics},
  publisher = {Cambridge University Press (CUP)},
  author = {Kaneelil,  Paul R. and Tojo,  Kazuki and Farsoiya,  Palas Kumar and Deike,  Luc and Stone,  Howard A.},
  year = {2026},
  month = jan 
}

@article{Kamani2021,
  title = {Unification of the Rheological Physics of Yield Stress Fluids},
  volume = {126},
  ISSN = {1079-7114},
  url = {http://dx.doi.org/10.1103/PhysRevLett.126.218002},
  DOI = {10.1103/physrevlett.126.218002},
  number = {21},
  journal = {Physical Review Letters},
  publisher = {American Physical Society (APS)},
  author = {Kamani,  Krutarth and Donley,  Gavin J. and Rogers,  Simon A.},
  year = {2021},
  month = may 
}

@article{Eggers2025,
  author = {Jens Eggers and James E. Sprittles and Jacco H. Snoeijer},
  title = {Coalescence Dynamics},
  journal = {Annual Review of Fluid Mechanics},
  year = {2025},
  volume = {57},
  pages = {TBD}
}

@article{Varma2022,
  author = {Sarath Chandra Varma and Debayan Dasgupta and Aloke Kumar},
  title = {Elasticity Can Affect Droplet Coalescence},
  journal = {Physics of Fluids},
  year = {2022},
  volume = {34},
  number = {1},
  pages = {013109},
  doi = {10.1063/5.0077144}
}

@article{Kern2022,
  author = {Vanessa R. Kern and Torstein Sæter and Andreas Carlson},
  title = {Viscoplastic Sessile Drop Coalescence},
  journal = {Physical Review Fluids},
  year = {2022},
  volume = {7},
  number = {9},
  pages = {093601},
  doi = {10.1103/PhysRevFluids.7.093601}
}

@article{Jalaal2021,
  author = {M. Jalaal and B. Stoeber and N. Balmforth},
  title = {Spreading of Viscoplastic Droplets},
  journal = {Journal of Fluid Mechanics},
  year = {2021},
  volume = {914},
  pages = {A21},
  doi = {10.1017/jfm.2021.101}
}

@article{Jalaal2015,
  author = {M. Jalaal and N. J. Balmforth and B. Stoeber},
  title = {Slip of Spreading Viscoplastic Droplets},
  journal = {Langmuir},
  year = {2015},
  volume = {31},
  number = {44},
  pages = {12071--12077},
  doi = {10.1021/acs.langmuir.5b02634}
}

@article{Jalaal2018,
  author = {M. Jalaal and C. Seyfert and B. Stoeber and N. J. Balmforth},
  title = {Gel-Controlled Droplet Spreading},
  journal = {Journal of Fluid Mechanics},
  year = {2018},
  volume = {837},
  pages = {115--137},
  doi = {10.1017/jfm.2017.821}
}

@article{vanderKolk2023,
  author = {J. van der Kolk and D. Tieman and M. Jalaal},
  title = {Viscoplastic Lines: Printing a Single Filament of Yield Stress Material on a Surface},
  journal = {Journal of Fluid Mechanics},
  year = {2023},
  volume = {958},
  pages = {A34},
  doi = {10.1017/jfm.2023.277}
}

@article{wei2023go,
  title={Go with the flow: Rheological requirements for direct ink write printability},
  author={Wei, Peiran and Cipriani, Ciera and Hsieh, Chia-Min and Kamani, Krutarth and Rogers, Simon and Pentzer, Emily},
  journal={Journal of Applied Physics},
  volume={134},
  number={10},
  year={2023},
  publisher={AIP Publishing}
}

@article{sauret2026fluid,
  title={Fluid Mechanics Challenges in Direct-Ink-Writing Additive Manufacturing},
  author={Sauret, Alban and Ray, Tyler R and Compton, Brett G},
  journal={Annual Review of Fluid Mechanics},
  volume={58},
  year={2026},
  publisher={Annual Reviews}
}

@article{saadi2022direct,
  title={Direct ink writing: a 3D printing technology for diverse materials},
  author={Saadi, MASR and Maguire, Alianna and Pottackal, Neethu T and Thakur, Md Shajedul Hoque and Ikram, Maruf Md and Hart, A John and Ajayan, Pulickel M and Rahman, Muhammad M},
  journal={Advanced Materials},
  volume={34},
  number={28},
  pages={2108855},
  year={2022},
  publisher={Wiley Online Library}
}

@article{martouzet2021dynamic,
  title={Dynamic arrest during the spreading of a yield stress fluid drop},
  author={Martouzet, Gr{\'e}goire and J{\o}rgensen, Loren and Pelet, Yoann and Biance, Anne-Laure and Barentin, Catherine},
  journal={Physical Review Fluids},
  volume={6},
  number={4},
  pages={044006},
  year={2021},
  publisher={APS}
}

@article{hinch2021oldroyd,
  title={Oldroyd B, and not A?},
  author={Hinch, John and Harlen, Oliver},
  journal={Journal of Non-Newtonian Fluid Mechanics},
  volume={298},
  pages={104668},
  year={2021},
  publisher={Elsevier}
}

@article{bouma2022optical,
  title={Optical coherence tomography},
  author={Bouma, Brett E and de Boer, Johannes F and Huang, David and Jang, Ik-Kyung and Yonetsu, Taishi and Leggett, Cadman L and Leitgeb, Rainer and Sampson, David D and Suter, Melissa and Vakoc, Ben J and others},
  journal={Nature Reviews Methods Primers},
  volume={2},
  number={1},
  pages={79},
  year={2022},
  publisher={Nature Publishing Group UK London}
}

@article{huang1991optical,
  title={Optical coherence tomography},
  author={Huang, David and Swanson, Eric A and Lin, Charles P and Schuman, Joel S and Stinson, William G and Chang, Warren and Hee, Michael R and Flotte, Thomas and Gregory, Kenton and Puliafito, Carmen A and others},
  journal={science},
  volume={254},
  number={5035},
  pages={1178--1181},
  year={1991},
  publisher={American Association for the Advancement of Science}
}

@article{DAngelo2022,
  title = {Spreading of droplets under various gravitational accelerations},
  volume = {93},
  ISSN = {1089-7623},
  url = {http://dx.doi.org/10.1063/5.0105624},
  DOI = {10.1063/5.0105624},
  number = {11},
  journal = {Review of Scientific Instruments},
  publisher = {AIP Publishing},
  author = {D’Angelo,  Olfa and Kuthe,  Felix and van Nieuwland,  Kasper and Ederveen Janssen,  Clint and Voigtmann,  Thomas and Jalaal,  Maziyar},
  year = {2022},
  month = nov 
}

@article{heitmeier2025spreading,
  title={Spreading droplets of yield-stress fluids with and without gravity},
  author={Heitmeier, Linnea and D'Angelo, Olfa and Jalaal, Maziyar and Voigtmann, Thomas},
  journal={arXiv preprint arXiv:2507.04379},
  year={2025}
}

@article{ho2025direct,
  title={Direct Ink Writing of Conductive Hydrogels},
  author={Ho, Monica and Ramirez, Aline Braz and Akbarnia, Negar and Croiset, Eric and Prince, Elisabeth and Fuller, Gerald G and Kamkar, Milad},
  journal={Advanced Functional Materials},
  volume={35},
  number={22},
  pages={2415507},
  year={2025},
  publisher={Wiley Online Library}
}

@article{colanges20232,
  title={2.5 D printing of a yield-stress fluid},
  author={Colanges, Simon and Tourvieille, Jean-No{\"e}l and Lidon, Pierre and Leng, Jacques},
  journal={Scientific Reports},
  volume={13},
  number={1},
  pages={5155},
  year={2023},
  publisher={Nature Publishing Group UK London}
}

@article{jang2021effect,
  title={Effect of material extrusion process parameters on filament geometry and inter-filament voids in as-fabricated high solids loaded polymer composites},
  author={Jang, Sungwoo and Boddorff, Andrew and Jang, Dong June and Lloyd, Jacob and Wagner, Karla and Thadhani, Naresh and Brettmann, Blair},
  journal={Additive Manufacturing},
  volume={47},
  pages={102313},
  year={2021},
  publisher={Elsevier}
}

@article{yoon2025minimizing,
  title={Minimizing defects in extrusion-based additive manufacturing through controlled filament spreading and overlapping},
  author={Yoon, Heedong and Yancheshme, Amir Azimi and Butler, Rhys and Palmese, Giuseppe R and Alvarez, Nicolas J},
  journal={Colloids and Surfaces A: Physicochemical and Engineering Aspects},
  pages={136879},
  year={2025},
  publisher={Elsevier}
}

@article{saengow2025stretching,
  title={Stretching the printability metric in direct-ink writing with highly extensible yield-stress fluids},
  author={Saengow, Chaimongkol and Sen, Samya and Yus, Joaquin and Lovrich, Eliza E and Hoika, Amanda G and Chang, Kelly M and Pfeil, Arielle A and Haug, Nellie and Johnson, Amy J Wagoner and Ewoldt, Randy H},
  journal={arXiv preprint arXiv:2501.12630},
  year={2025}
}

@article{friedrich2024simulated,
  title={Simulated inter-filament fusion in embedded 3D printing},
  author={Friedrich, Leanne M and Gunther, Ross T},
  journal={Biofabrication},
  volume={17},
  number={1},
  pages={015022},
  year={2024},
  publisher={IOP Publishing}
}

@article{grosskopf2018viscoplastic,
  title={Viscoplastic matrix materials for embedded 3D printing},
  author={Grosskopf, Abigail K and Truby, Ryan L and Kim, Hyoungsoo and Perazzo, Antonio and Lewis, Jennifer A and Stone, Howard A},
  journal={ACS applied materials \& interfaces},
  volume={10},
  number={27},
  pages={23353--23361},
  year={2018},
  publisher={ACS Publications}
}

@article{kamani2024brittle,
  title={Brittle and ductile yielding in soft materials},
  author={Kamani, Krutarth M and Rogers, Simon A},
  journal={Proceedings of the National Academy of Sciences},
  volume={121},
  number={22},
  pages={e2401409121},
  year={2024},
  publisher={National Academy of Sciences}
}

@article{pagani2024no,
  title={No yield stress required: Stress-activated flow in simple yield-stress fluids},
  journal={Journal of Rheology},
  volume={68},
  number={2},
  pages={155--170},
  year={2024},
  publisher={AIP Publishing}
}

@article{divoux2024ductile,
  title={Ductile-to-brittle transition and yielding in soft amorphous materials: perspectives and open questions},
  author={Divoux, Thibaut and Agoritsas, Elisabeth and Aime, Stefano and Barentin, Catherine and Barrat, Jean-Louis and Benzi, Roberto and Berthier, Ludovic and Bi, Dapeng and Biroli, Giulio and Bonn, Daniel and others},
  journal={Soft Matter},
  volume={20},
  number={35},
  pages={6868--6888},
  year={2024},
  publisher={Royal Society of Chemistry}
}

@article{saramito2021new,
  title={A new brittle-elastoviscoplastic fluid based on the Drucker--Prager plasticity},
  author={Saramito, Pierre},
  journal={Journal of Non-Newtonian Fluid Mechanics},
  volume={294},
  pages={104584},
  year={2021},
  publisher={Elsevier}
}

@article{larson2015constitutive,
  title={Constitutive equations for thixotropic fluids},
  author={Larson, RG},
  journal={Journal of Rheology},
  volume={59},
  number={3},
  pages={595--611},
  year={2015},
  publisher={AIP Publishing}
}

@article{balmforth2014yielding,
  title={Yielding to stress: recent developments in viscoplastic fluid mechanics},
  author={Balmforth, Neil J and Frigaard, Ian A and Ovarlez, Guillaume},
  journal={Annual review of fluid mechanics},
  volume={46},
  number={1},
  pages={121--146},
  year={2014},
  publisher={Annual Reviews}
}

@article{chaparian2019adaptive,
  title={An adaptive finite element method for elastoviscoplastic fluid flows},
  author={Chaparian, Emad and Tammisola, Outi},
  journal={Journal of Non-Newtonian Fluid Mechanics},
  volume={271},
  pages={104148},
  year={2019},
  publisher={Elsevier}
}

% \clearpage
% \renewcommand\thefigure{S\arabic{figure}}
% \setcounter{figure}{0}    
% \section*{Supplementary Information}
% \subsection*{Supplementary Movies}
% Supplementary ...\\

\pagebreak

\appendix
\section*{Appendices}

\section{Rheology of samples}
\label{section:app:rheo}
\begin{figure}[htbp]
\centering
\includegraphics[width=1\textwidth]{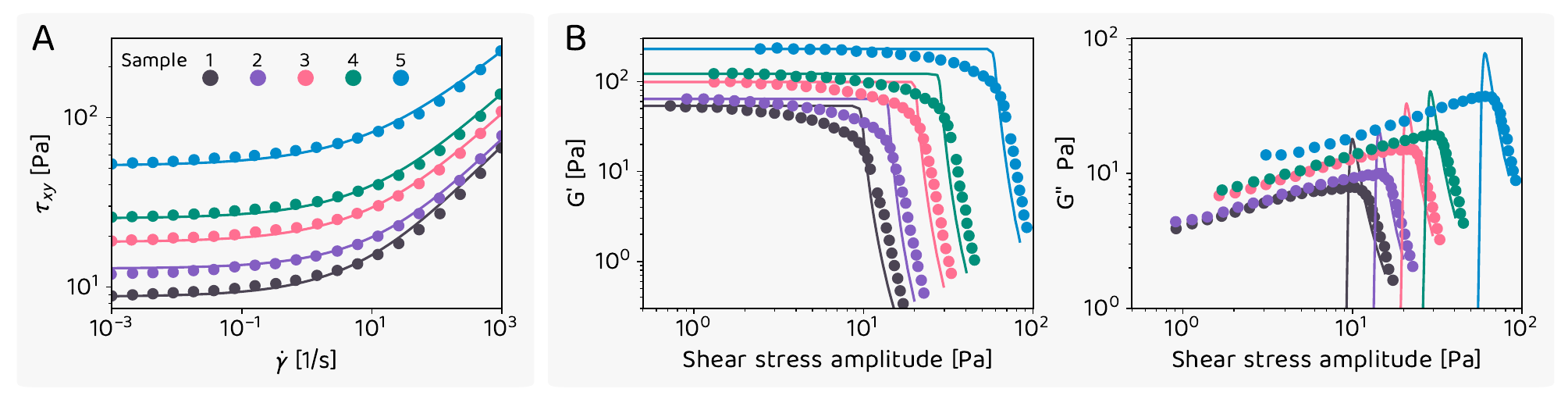}
\caption{Rheology data from simple shear rheometer measurements (circles) and from the fitted Saramito model (lines). Panel A: steady-state shear stress as a function of applied shear-rate, obtained from steady shear measurements. Panel B: elastic ($G'$) and loss ($G''$) modulus as a function of shear stress amplitude, obtained from oscillatory shear tests by performing a sweep in shear amplitude. The parameters used for the Saramito model curves can be seen in table \ref{table:rheology_parameters}.}
\label{fig:rheology}
\end{figure}

Rheological assessments were conducted using an Anton Paar MCR 302 rheometer, employing a cone-and-plate geometry with a 1 degree cone angle. Samples were place on the flat, stationary plate surface, which was coated with rough sandpaper to establish a no-slip boundary condition. 
Steady shear experiments were performed by executing shear rate sweeps in the range [0.01, 1000] $s^{-1}$. 
% The rheometer automatically adjusted the duration of each shear rate step within a range of 15 to 30 seconds, ensuring that a steady state was attained for each measurement. 
High-to-low rate sweeps were chosen to avoid short-term transient effects.
% associated with the transition from viscoelastic solid to mobile liquid during fluidization near the yield point. 
The flow curves obtained from these experiments, shown in Fig.~\ref{fig:rheology}A, were used to fit the parameters of the Saramito model (lines), which provides good agreement with the steady-state shear stress.

Oscillatory shear tests were also performed with the same rheometer setup. An oscillatory strain with the form $\gamma = \gamma_0\sin{\left(\omega\,t\right)}$ was applied to the material. The oscillation frequency was kept fixed at $\omega = 1$\,rad/s, while the strain amplitude $\gamma_0$ varied in the range [0.1, 100]. The storage $(G')$  and loss $(G'')$ moduli obtained from these sweeps are presented in figure~\ref{fig:rheology}B. The low-amplitude plateau values of $G'$ were used as the elastic modulus $G$ for the Saramito model. The solvent viscosity $\eta_s$ is kept fixed as the viscosity of water. Lines in figure~\ref{fig:rheology}B show this fit of the Saramito model. The $G'$  curve shows a reasonably good agreement with experiments, with some significant discrepancy being present only around the yielding transition, since the Saramito model presents a non-smooth $G'$ curve at the yield-stress. For the $G''$ curve the discrepancies are more pronounced. Not only the Saramito model fails to predict the transition region accurately, but the entire region of $G''$ below the yield stress is severely underpredicted since in its unyielded phase, the Saramito model presents only solvent viscosity, which leads to a constant and small $G''$, since we take the viscosity of water as our $\eta_s$. This unphysical prediction of the unyielded $G''$ is a well-known characteristic of the Saramito model and has recently been tackled by unified EVP models such as in \cite{Kamani2021,kamani2024brittle}. All the fitted parameters for each of the five materials used in this article are presented in table \ref{table:rheology_parameters}. We note once again that the values of yield-stress shown in table \ref{table:rheology_parameters} are obtained assuming the matrix norm $\lVert \cdot \rVert$ defined in section \ref{sec:formulations}, which does not include a factor of $\sqrt{2}$. For authors that do use this factor in their definition, the yield-stress value should also be corrected by this factor.

\begin{table}[h!]
\centering
\caption{Combined rheological and dimensionless parameters. For each sample, the area of the printed filaments is processed from the cross-section images and the length-scale $\mathcal{L}$ (where $\mathcal{L}^2$ is the combined area of both filaments) is used in the dimensionless groups. This length-scale slightly varies between experiments, but is always close to $\mathcal{L} = 1.2 \times10^{-3}$m.}
\begin{tabular}{
c
S[table-format=2.2]
S[table-format=3.2]
S[table-format=1.2]
S[table-format=1.2]
S[table-format=1.2]
S[table-format=1.2]
S[table-format=1.2]
S[table-format=1.2]
S[table-format=1.2]
}
\toprule
& \multicolumn{5}{c}{\textbf{Rheological parameters}} & \multicolumn{4}{c}{\textbf{Dimensionless groups}} \\
\cmidrule(lr){2-6} \cmidrule(lr){7-10}
{Sample} &
{$\tau_y$ [Pa]} &
{$G$ [Pa]} &
{$K$ [Pa\,s$^n$]} &
{$\eta_s$ [Pa\,s]} &
{$n$} &
{$Oh_s$} &
{$Oh_p$} &
{$\mathcal{J}$} &
{$De$} \\
\midrule
1 & 12.56 & 53.65 & 2.50 & 0.001 & 0.52 & 0.0034 & 0.665 & 0.21 & 0.728 \\
2 & 18.55 & 64.00 & 2.60 & 0.001 & 0.52 & 0.0033 & 0.695 & 0.32 & 0.624 \\
3 & 26.50 & 98.80 & 4.50 & 0.001 & 0.49 & 0.0034 & 1.016 & 0.44 & 0.617 \\
4 & 36.68 & 122.12 & 6.00 & 0.001 & 0.49 & 0.0034 & 1.354 & 0.61 & 0.666 \\
5 & 75.40 & 230.31 & 14.00 & 0.001 & 0.45 & 0.0034 & 2.575 & 1.29 & 0.655 \\
\bottomrule
\end{tabular}
\label{table:rheology_parameters}
\end{table}

\section{Small elasticity as regularization for the VP limit}
\label{section:regularization}

In this appendix we describe the regularization strategy used in section~\ref{sec:purely_vp} to obtain results in the purely viscoplastic Herschel--Bulkley limit. This limit is known to be computationally challenging due to the diverging apparent viscosity that appears at small shear rates~\cite{balmforth2014yielding}.

This challenge can be alleviated by using the Saramito model with a very small (but non-zero) relaxation time. This allows us to view the relaxation time as a numerical parameter that can be tuned to obtain the desired trade-off between numerical stability and reproducibility of the pure VP limit.

The numerical issue in pure Herschel-Bulkley rheology is associated with an apparent viscosity $\eta_{HB}$ that diverges for small shear rates:
\begin{equation}
\label{eq:hb_pure}
\eta_{\mathrm{HB}}( \dot{\boldsymbol{\gamma}}) = K \, \lVert \dot{\boldsymbol{\gamma}}\rVert^{\,n-1} + \frac{\tau_y}{\lVert \dot{\boldsymbol{\gamma}}\rVert},
\end{equation}
where the stress tensor is then defined in a generalized Newtonian form as $\boldsymbol{\tau} = \eta_{\mathrm{HB}} \dot{ \boldsymbol{\gamma} }$. In literature, the divergence of viscosity in HB models is often remedied by sharply truncating its maximum value with a given constant \citep{Sanjay2021} or by using the smooth non-diverging approximation proposed by \citet{Papanastasiou1987}:
\begin{equation}
\label{eq:papanastasiou}
\eta_{\mathrm{PP}}( \dot{\boldsymbol{\gamma}}) = \left( K \, \lVert \dot{\boldsymbol{\gamma}}\rVert^{\,n} + \tau_y \right) \frac{ \left( 1 - e^{-\lVert \dot{\boldsymbol{\gamma}}\rVert/\epsilon} \right)}{\lVert \dot{\boldsymbol{\gamma}}\rVert}.
\end{equation}

The main disadvantages of this approach are related to the lack of direct physical interpretation given to the parameter $\epsilon$ and the fact that, with a truncated maximum viscosity, the true solid regime will never be achieved, but only a highly viscous fluid. On the other hand, it presents the advantage of not introducing any new internal timescale to the constitutive equation. The material still has a truly instantaneous response to shear rate. 

By using the Saramito model with a small relaxation time $\lambda$, we view $\lambda$ as a numerical parameter that has a clear physical meaning, since it represents the needed time for the internal stresses of the material to develop and relax. 

To illustrate the practical effects of the relaxation time as VP regularizator, we will perform a simple transient startup flow. Consider a 2D geometry with height $2H$ in which a constant shear rate $\dot{\gamma}_{xy}$ is imposed. By using $H$ as characteristic length, $H\dot{\gamma}_{xy}$ as characteristic velocity and $\eta_p\dot{\gamma}_{xy}$ as characteristic polymeric stress, the Saramito equation can be written as
\begin{equation}
\mathrm{Wi} \, \overset{\nabla}{\bar{\boldsymbol{\tau}}^p}
+ \max\!\left[0, 
\left(\frac{|\bar{\dot{\boldsymbol{\gamma}}}| - \mathrm{Bi}}
{ |\bar{\dot{\boldsymbol{\gamma}}}|}\right)^{1/n}\right]
\bar{\boldsymbol{\tau}}^p
 = \bar{\dot{\boldsymbol{\gamma}}},
\label{eq:saramito_shear}
\end{equation}
where $\mathrm{Bi} = \tau_y/(K\dot{\gamma}_{xy}^n)$ is the Bingham number and $\mathrm{Wi} = (K\dot{\gamma}_{xy}^n)/G$ is the Weissenberg number, which takes the form of a nondimensional relaxation time and will be used as our viscoplastic regularization parameter in this section.

\begin{figure}[htbp]
\centering
\includegraphics[width=1\textwidth]{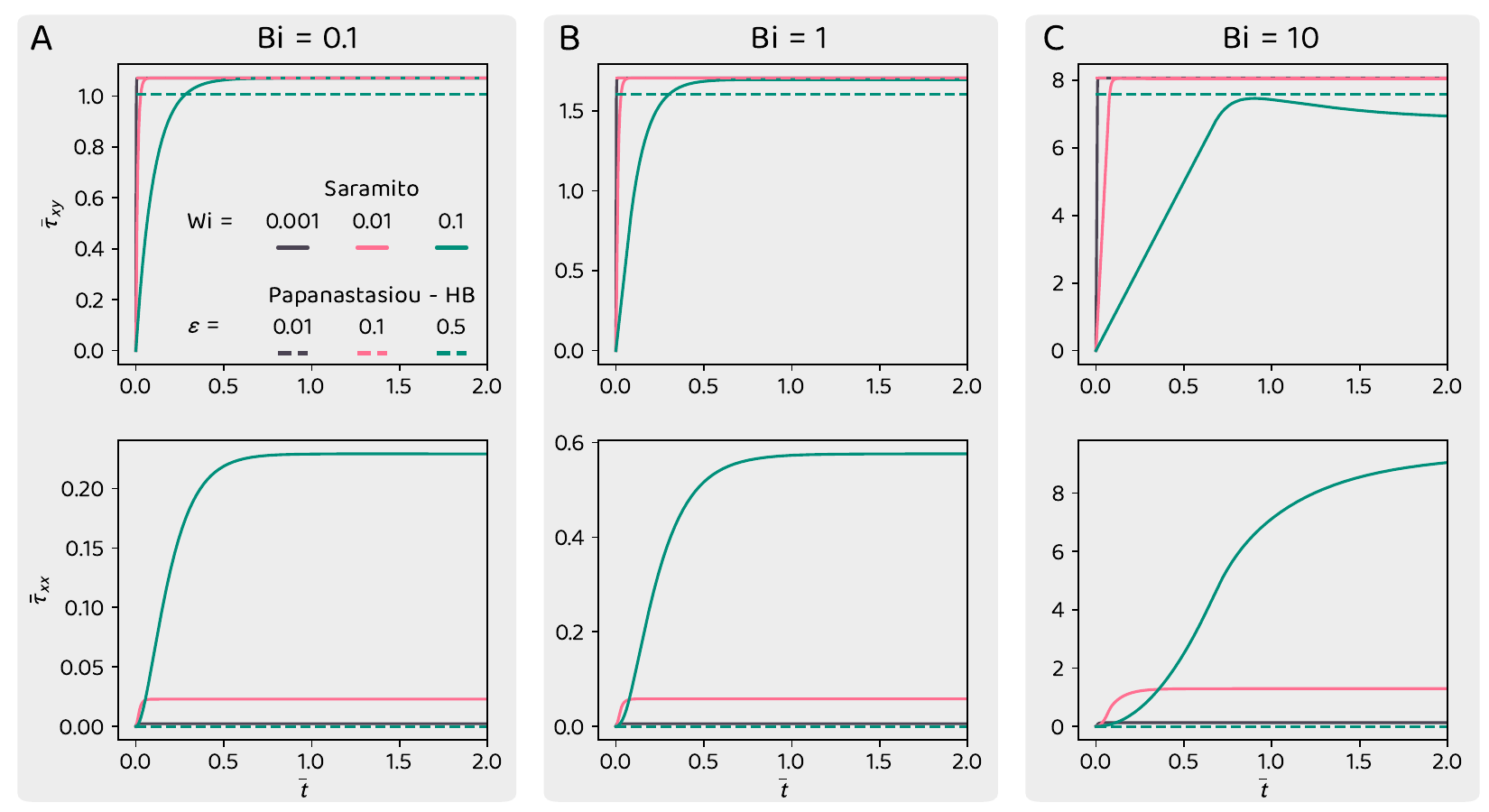}
\caption{Time evolution of polymeric stress tensor during a steady shear flow. The Saramito model presents a small transient period and creates some normal stress, both of these outcomes are minimal with $\mathrm{Wi} = 10^{-3}$ and the curves are almost indistinguishable from those of the Herschel-Bulkley model. }
\label{fig:regularization_transient}
\end{figure}

Figure \ref{fig:regularization_transient} shows the transient stress response of materials with three different Bingham numbers. For each Bingham number, different amounts of regularization are compared with each other. In a true Herschel-Bulkley material, the stress would instantaneously respond to a given shear rate, such that only a horizontal line should be expected in these plots. This behaviour is correctly captured by the Papanastasiou regularization, at the expense of the predicted stress being lower than expected, depending on the regularization parameter $\epsilon$. For the Saramito model, a finite transient period is introduced with the small value of $\mathrm{Wi}$. We see, however, that this period is very small for $\mathrm{Wi} \leq 10^{-2}$, such that the material can be considered to have a quasi-instantaneous response. The introduction of $\mathrm{Wi}$ also slightly alters the steady-state shear and normal stresses from the expected pure VP, but we see from these plots that this alteration is very small for $\mathrm{Wi} = 10^{-3}$.

\begin{figure}[htbp]
\centering
\includegraphics[width=0.6\textwidth]{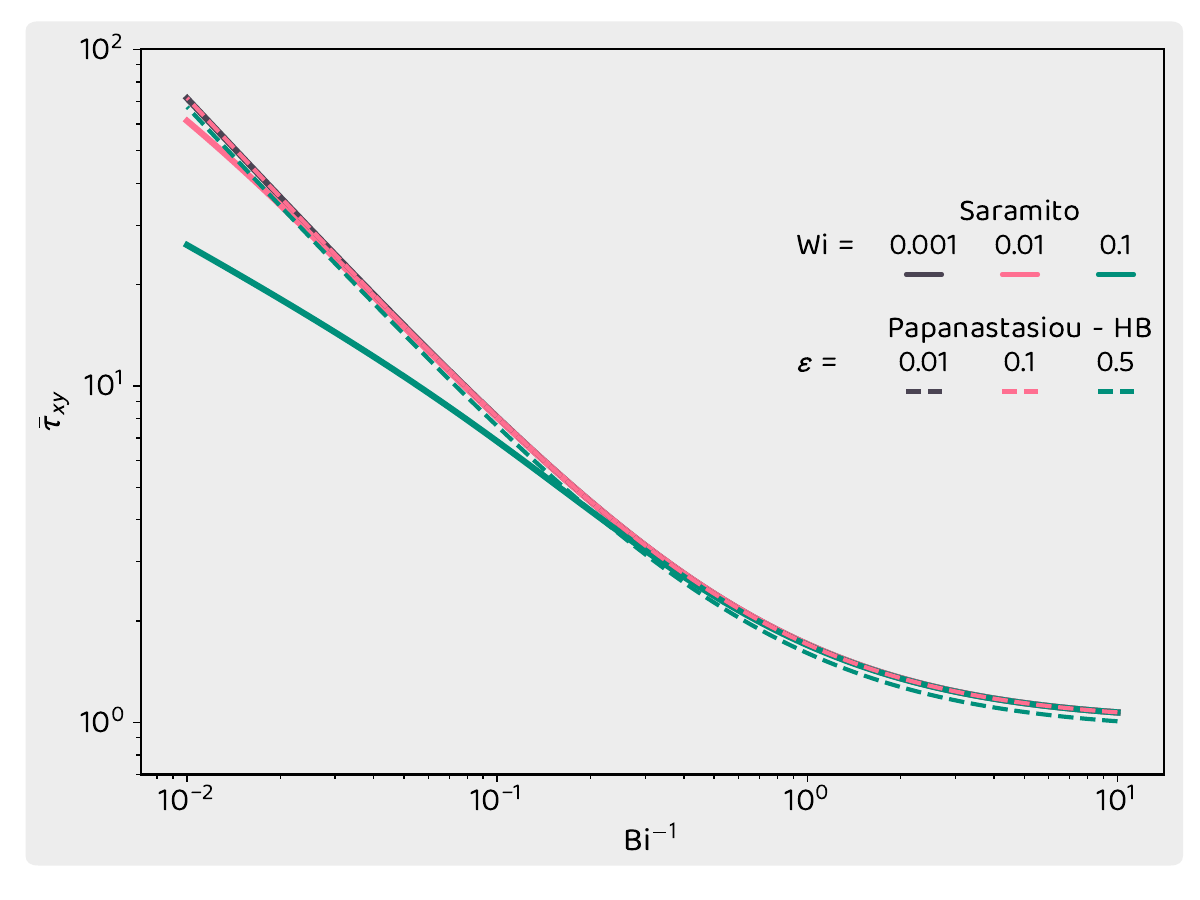}
\caption{Nondimensional flow curves for the Saramito and Herschel-Bulkley models. For $\mathrm{Wi} = 10^{-3}$, the two methods are almost indistinguishable.}
\label{fig:regularization_steady}
\end{figure}

To further investigate the effect of a small Weissenberg number in the steady-state stress, we show in Figure \ref{fig:regularization_steady} a sweep in $\mathrm{Bi}^{-1}$, which can be seen as a nondimensional equivalent of a shear rate sweep. We see that for small shear rates (high Bingham number) the results differ the most from the pure Herschel-Bulkley limit, since this is the area in which diverging viscosities tend to appear. However, we see again that for $\mathrm{Wi} = 10^{-3}$ the Saramito flow curve is almost indistinguishable from the pure Herschel-Bulkley, which validates the useage of the Weissenberg number as a regularization parameter for the purely viscoplastic limit.

While effective for the present purposes, the regularization strategy adopted here warrants further systematic investigation, particularly for non-viscometric and strongly transient flows and a more detailed assessment of how elasto-plastic effects influence both regularization and macroscopic dynamics~\cite{chaparian2019adaptive}

\section{Effect of the initial shape}
\label{sec:shape_dependence}

\begin{figure}[t!]
\centering
\includegraphics[width=1\textwidth]{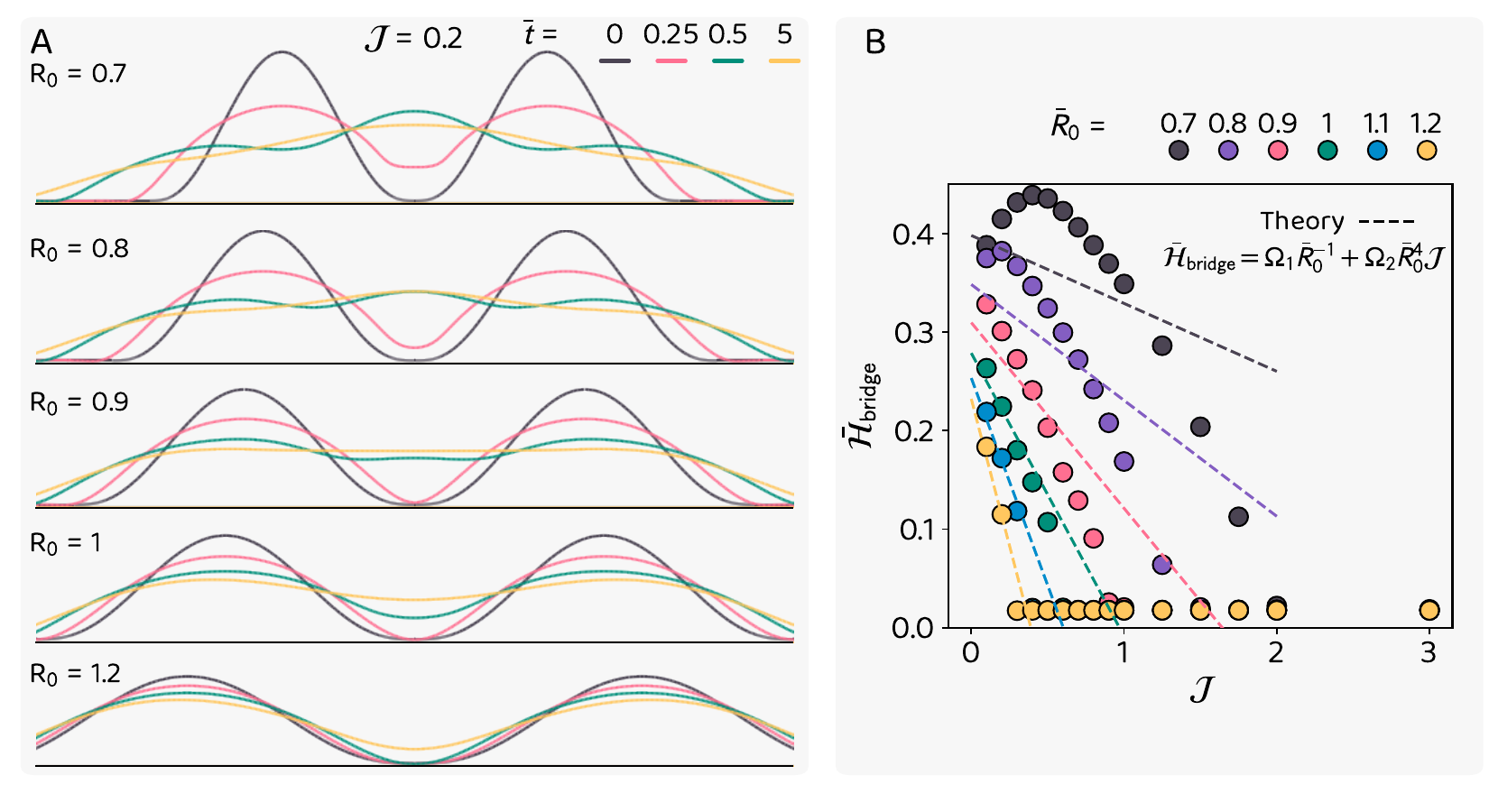}
\caption{Effect of initial shape on coalescence outcome, studied by varying the base radius $\bar{R}_0$ in the initial condition~\eqref{eq:initial_shape}. Panel A: snapshots of filaments over time for different $\bar{R}_0$. Narrow filaments present high curvatures, leading to stronger coalescence. Panel B: final bridge height as a function of $\mathcal{J}$ for multiple $\bar{R}_0$. We keep fixed $\mathrm{Oh}_s = 1/90$, $\mathrm{Oh}_p = 8/90$, $n = 0.5$, and $\mathrm{De} = 10^{-3}$; the small Deborah number ensures that elastic effects are negligible. An animation of the simulations in this figure can be seen in supplementary video IV.}
\label{fig:shape_dependence}
\end{figure}

In this appendix, we illustrate how the initial condition can strongly influence the final shape after coalescence. All simulations are performed in the quasi-viscoplastic regime defined in Sec.~\ref{sec:purely_vp}, with fixed parameters $\mathrm{Oh}_s = 1/90$, $\mathrm{Oh}_p = 8/90$, $n = 0.5$, and $\mathrm{De} = 10^{-3}$; the small Deborah number ensures that elastic effects are negligible. Because the initial condition can be modified in several ways, we choose to vary the base radius $\bar{R}_0$ in Eq.~\eqref{eq:initial_shape}; see Fig.~\ref{fig:shape_dependence}. In panel A, we show snapshots from simulations with $\mathcal{J}=0.2$ for different values of $\bar{R}_0$. These can also be dynamically visualized in supplementary video IV. Since our nondimensionalization keeps the filament cross-sectional area constant, decreasing $\bar{R}_0$ necessarily increases the initial filament height. Taller filaments therefore exhibit a larger effective curvature at the symmetry point between the two filaments. This increased curvature generates a stronger capillary pressure, which can overcome higher yield stresses (encoded in $\mathcal{J}$). At a critical plastocapillary number $\mathcal{J}_c$, however, the capillary pressure induced by the initial condition is insufficient to initiate flow, and the bridge fails to grow. This behavior is evident in panel B, where the final bridge height $\bar{\mathcal{H}}_{\mathrm{bridge}}$ is shown to persist to significantly larger values of $\mathcal{J}$ as $\bar{R}_0$ decreases, confirming that increased initial curvature promotes coalescence at higher yield stress.

In panel B of Fig.~\ref{fig:shape_dependence}, we also include the theoretical prediction from Eq.~\eqref{eq:theory_scaling_1}. Using constant prefactors $\Omega_1 = 0.28$ and $\Omega_2 = -0.29$ we find good agreement between the scaling law and the simulation results for shallow filaments ($\bar{R}_0>1$). We also note that the value of $\Omega_1$ is very close to our theoretical approximation $\Omega_1 = 1/4$ made in section~\ref{sec:theory}. For steep filaments, however, this simple scaling breaks down, as the prefactors $\Omega_1$ and $\Omega_2$ likely depend on geometry and can no longer be treated as constants. Additionally, for $\bar{R}_0 = 0.7$, the final bridge height becomes non-monotonic at small values of $\mathcal{J}$. In this regime, the large initial capillary pressure drives rapid bridge growth, leading to an inertial overshoot beyond the eventual steady height (see supplementary video IV). For small $\mathcal{J}$, surface tension subsequently flattens the resulting central peak. For intermediate values of $\mathcal{J}$, however, surface tension is insufficient to fully relax this peak, which becomes frozen into the final shape. At larger $\mathcal{J}$, the bridge arrests earlier and the peak never forms. This arrested peak at intermediate $\mathcal{J}$ gives rise to the non-monotonic dependence observed for $\bar{R}_0=0.7$.

In the context of DIW, these results highlight that the effective initial condition is tightly coupled to the deposition process itself. Parameters such as nozzle height and the rheological history of the ink inside the nozzle directly determine the initial filament geometry and curvature. As a result, controlling deposition conditions is essential for achieving predictable filament coalescence and controlling residual surface topography in printed structures.

\end{document}